%% file: main.tex
\documentclass[manuscript]{acmart} 
\AtBeginDocument{%
  }

\setcopyright{acmlicensed}
\copyrightyear{2025}
\acmYear{2025}
\acmDOI{XXXXXXX.XXXXXXX}

\usepackage[showcomments,showtodos]{labsedc}
\usepackage{extracommands}

\usepackage{hyperref}
\usepackage{balance}
\usepackage{placeins} 
\usepackage{array} 
\usepackage{multirow} 
\usepackage{adjustbox} 
\usepackage[normalem]{ulem}
\usepackage[framemethod=TikZ]{mdframed}
\newmdenv[
  skipabove=8pt,
  skipbelow=8pt,
  leftmargin=0pt,
  rightmargin=0pt,
  innertopmargin=6pt,
  innerbottommargin=6pt,
  linecolor=black,
  linewidth=0.5pt,
  roundcorner=4pt,
  backgroundcolor=gray!10
]{acmmdframed}

\begin{document}

\title{Large Language Models  for Software Testing: A Research Roadmap}


\author{Cristian Augusto}
\affiliation{%
  \institution{University of Oviedo}
  \city{Gijón}
  \country{Spain}}
\email{augustocristian@uniovi.es}

\author{Antonia Bertolino}
  \email{antonia.bertolino@isti.cnr.it}
\affiliation{%
  \institution{GSSI}
  \city{L'Aquila}
  \country{Italy}
}

\author{Guglielmo De Angelis}
\affiliation{%
 \institution{IASI-CNR}
 \city{Rome} 
 \country{Italy}}
\email{guglielmo.deangelis@iasi.cnr.it}

\author{Francesca Lonetti}
\affiliation{%
  \institution{ISTI-CNR}
  \city{Pisa}
  \country{Italy}}
  \email{francesca.lonetti@isti.cnr.it}

\author{Jesús Morán}
\affiliation{%
    \institution{University of Oviedo}
  \city{Gijón}
  \country{Spain}}
\email{moranjesus@uniovi.es}

\renewcommand{\shortauthors}{Augusto et al.}

\begin{abstract}
Large Language Models (LLMs) are starting to be profiled as one of the most significant disruptions in the Software Testing field. 
Specifically, they have been successfully applied in software testing tasks such as generating test code, or summarizing documentation. 
This potential has attracted hundreds of researchers, resulting in dozens of new contributions every month, hardening researchers to stay at the forefront of the wave. Still, to the best of our knowledge, no prior work has provided a structured vision of the progress and most relevant research trends in LLM-based testing.
In this article, we aim to provide a roadmap that illustrates its current state, grouping the contributions into different categories, and also sketching the most promising and active research directions for the field. 
To achieve this objective, we have conducted a semi-systematic literature review, collecting articles and mapping them into the most prominent categories, reviewing the current and ongoing status, and analyzing the open challenges of LLM-based software testing. Lastly, we have outlined several expected long-term impacts of LLMs over the whole software testing field.

\end{abstract}

\ccsdesc[500]{General and reference~Surveys and overviews}
\ccsdesc[500]{General and reference~Validation}
\ccsdesc[500]{General and reference~Verification}
\ccsdesc[500]{Computing methodologies~Natural language processing}
\ccsdesc[500]{Software and its engineering~Software testing and debugging}
\keywords{Software Testing, Large Language Model, Roadmap}


\maketitle

\newcommand{\numarticles}{130 }
\newcommand{\numunit}{40 }
\newcommand{\numhighend}{28 }
\newcommand{\numoracle}{26 }
\newcommand{\numreflections}{23 }
\newcommand{\numaugment}{12 }
\newcommand{\numagents}{8 }
\newcommand{\numnonfuct}{8 }

\input{01_intro}
\input{02_relatedwork}

\input{03_mapping}

\input{04_papercomporae}
\input{05_llmUsage}
\input{06_reflection}
\input{07_conclusion}

\begin{acks}

The work was partially supported by \matisseproject, in part by  \equavelproject, and also by \fairproject. 

\end{acks}

\bibliographystyle{ACM-Reference-Format}
\bibliography{biblio}

\end{document}

%% file: 01_intro.tex
\section{Introduction}
\label{intro}

Transformer-based large language models (LLMs) have revolutionized the Natural Language Processing (NLP) field and, in turn, have become the cornerstone of technological progress in a broad range of fields, even challenging the current state-of-the-art tests for intelligence~\cite{Savage2024}.
LLMs are neural architectures that mimic human reasoning skills by learning patterns from vast human-crafted text corpora.
LLMs have shown their potential in various tasks, such as generating narratives and detailed answers to human questions, or correcting and improving grammar, or summarizing and easing the understanding of complex information.

In the Software Engineering (SE)  field~\cite{Ozkaya2023}, LLMs are reshaping how repetitive and low-value tasks (also known as TOIL)~\cite{Betsy2016} are managed:
LLMs are being integrated in most IDEs (e.g., IntelliJ\footnote{https://www.jetbrains.com/idea/}, VSCode\footnote{https://code.visualstudio.com/}), providing tools that help developers in code synthesis~\cite{Fakhoury2024}, explain the code functionality~\cite{Nam2024}, detect software bugs and provide their fixes~\cite{Liu2024, Sobania2024}, as well as assist crafting the documentations and reports~\cite{Jin2023}. 
First reports from big tech companies like AWS~\cite{Omidvar2024} and Meta~\cite{Alshahwan2024} begin to realize the advantages of LLM-driven automation, as a support and complement of human engineers.

Software testing is a fundamental element of SE, representing between 15\% to 80\% of the total project budget~\cite{Polo2013, Xie2018}. This has generated increasing interest~\cite{Hou2024} in using LLMs in different tasks and levels of software testing.
Recent studies have shown the LLMs potential to speed up and automate different tasks like oracle generation~\cite{Molina2024}, automatic program repair~\cite{Plein2024}, test suite augmentation~\cite{Alshahwan2024}, test input generation~\cite{Jiang2024}, and both unit~\cite{ChenFSE2024} and system~\cite{Augusto2025} test cases generation.

Research on employing LLMs in software testing (hereafter referred to as ``LLM-based testing'') has been very active in last years, with dozens of new articles published monthly, or even weekly. In such a quickly evolving landscape, it is difficult for interested researchers to follow the progress and achievements of research, and especially to grasp a structured vision of the most relevant research trends. 
In this article, we aim to provide a roadmap that illustrates the current state of LLM-based testing research, organized into a set of article categories as they emerged from our study of literature; we also aim to sketching the most promising future research directions. While to address these aims we have conducted an extensive study of literature, we underscore that this work is not meant strictly as a systematic literature review (SLR), which is beyond the scope of our study: as we better explain in Section~\ref{sec:mapping}, we currently consider a roadmap as more practical and useful than a strict SLR.

The general objective of this work is divided into the following sub-objectives (SOs):

\begin{itemize}
    \item [SO1:] Derive a structured map of the most relevant research categories for LLM-based testing;
    \item [SO2:] Overview of the ongoing status of the LLM-based testing research along the identified categories;
    \item [SO3:] Analyze the most prominent open challenges projected into the depicted roadmap;
    \item [SO4:] Extrapolate the foreseen long-term impact of LLMs over the whole software testing field.
\end{itemize}

The remainder of this article is structured as follows: Section~\ref{sec:relatedwork} reviews the related work. Section~\ref{sec:mapping} outlines the process for retrieving literature. Section~\ref{sec:corpora} presents the retrieved articles and describes how they were classified to address SO1. Section \ref{sec:llmUsage} presents how LLMs are used in the different software testing categories considered, using a separate subsection for each category (SO2). Section~\ref{sec:reflections} analyzes the relevant LLM-based testing challenges and outlines the future research directions (SO3), and attempts a forecast of how this new research wave is going to impact the testing research (SO4). Finally, Section~\ref{sec:conclusion} presents the article's conclusions.

%% file: 02_relatedwork.tex
\section{Related Work}
\label{sec:relatedwork}

As said, the field of LLM-based testing has been very active in recent years, and several studies have already tried to review existing LLM-based testing techniques and tools, or to analyze LLMs' performances and interaction methods, as well as to synthesize research challenges and perspectives.

At time of writing, the work in~\cite{WangTSE2024} represents the most recent systematic survey on LLM-based testing. It covers 102 studies from 2019 to 2023  that are classified according to the software testing tasks for which LLMs are employed, the used LLM model, the type of prompt engineering, and LLM input. Also, challenges and potential opportunities are presented. 
Another survey on LLM-based testing is presented in~\cite{Fei2024}. It first presents the evolution of LLMs development from 2003 to 2021 and then analyzes 19 studies using LLM to optimize testing techniques that are classified according to two dimensions: how to generate automated test code efficiently and how to generate diverse test inputs.
Finally, the authors of~\cite{Santos2024}  rely on a questionnaire-based survey to collect quantitative data about the practical usage of LLMs in the various testing activities.
Our article differs from the previous ones since we do not aim to provide yet another systematic survey of LLM-based testing literature. Our main goal is to build a roadmap outlining the current status of the LLM-based testing field, together with a structured overview of the research directions and the long-term perspectives in the field. 
We followed a semi-systematic review approach~\cite{Snyder2019} that is a still rigorous but less comprehensive review, which allows us to identify and classify approaches on LLM-based testing as well as spot knowledge gaps within the literature.

Other reviews focus on using LLMs for test oracle automation. In particular, the work in~\cite{Molina2024} provides a roadmap covering the common types of automatically inferred oracles and discussing both the potential and threats of using LLMs for oracle automation. Similar works address the problem of the evaluation of LLM-based oracle generation models~\cite{Soneya2023}, their performances~\cite{ZhangTSE2025}, and evaluation metrics~\cite{Jiho2024, Zhongxin2023}.
Oracle generation is one of the categories we used for classifying articles in our roadmap, and we considered the above articles in the reflection category related to oracle generation of our roadmap.

Studies with a broader focus address the more general problem related to the adoption of LLMs in software engineering and also in software testing. For instance, the authors of~\cite{Hou2024} present a systematic survey aiming to understand how LLMs can be exploited in the different software engineering tasks, including software quality assurance and test generation, that are addressed in our work. 
Similarly, the authors of~\cite{Zheng2025}  present a general and recent systematic review that addresses the role of LLMs in the different software engineering activities, including code evaluation. It analyzes studies related to  LLM-based test case generation, new testing frameworks, and LLM-testing models, as in our work. 
The survey in~\cite{fan2023large} reviews the existing works on LLMs for software engineering activities, including software testing. It addresses test generation, test adequacy evaluation, test minimization, and test output prediction, also highlighting open problems and challenges in the field. 
The work in~\cite{Braberman2024} presents a systematic taxonomy of software engineering problems, including those related to software testing and a mapping of these problems to the  LLM-based approaches and tools that are used to solve them. Finally, the results of a questionnaire survey involving 100 software engineers on the adoption of generative AI in software development are presented in~\cite{Russo2024}. This work discusses software developers' perceptions about relevant aspects of the adoption of LLMs in software engineering, such as perceived usefulness, complexity, technological advantages, social influence, environmental factors, and others.  
Different from these surveys that span the whole software engineering aspects, we focus on software testing and deeply investigate relevant studies on LLM-based software testing.

Recent research has increasingly focused on evaluating LLM-driven testing processes, particularly in the context of unit testing, by proposing frameworks and pipelines for the assessment and analysis of the effectiveness of LLM-based testing as well as practical guidelines and recommendations for improving test results~\cite{OuedraogoArxiv2024, Li2025, Shang2024, Abdullin2025, Siddiq2024}.
Specifically, the work in~\cite{Li2025} introduces a benchmarking framework designed to compare the capabilities of LLMs with manual testing across multiple dimensions, including test case generation, bug detection, and error tracing, using open-source software projects as a reference context. The authors of~\cite{Shang2024}  provide a comprehensive empirical evaluation of 37 widely adopted LLMs, discussing critical issues related to the evaluation of LLMs in the software engineering community, such as data leakage,  fault-detection capability of generated test cases, and the comparison and selection of the evaluation metrics. 
Similarly, the authors of~\cite{OuedraogoArxiv2024}  evaluate the effectiveness of different models, such as GPT and Mistral, in generating unit test cases and how their performance is affected by different prompt engineering techniques.
The authors of~\cite{Abdullin2025} investigate the comparative performance of LLM-based testing and traditional automated test generation techniques, such as Search-Based Software Testing (SBST) and symbolic execution, highlighting relative strengths and limitations.
Finally, other works~\cite{Mathews2024, Boukhlif2024} classify the most used LLMs in software testing, according to: their input and output, the automated software testing types they perform, their capability to produce tests able to effectively identify bugs, and the application domains.
Unit test generation represents one of the main research categories of our roadmap, while critical challenges in LLM-based testing, such as prompt engineering, data leakage, and evaluation metrics, are examined in our work.
In this extended panorama, the original contribution of this study relies into our aim of providing a structured map of ongoing research, which can be used to better understand and position articles on the topic, usable even for framing research works appearing in future; at the same time, we propose the most relevant open problems for future research as emerging from the study of the literature trends.

%% file: 03_mapping.tex
\section{Literature mapping}
\label{sec:mapping}

In this work, we aim to sketch the main directions being explored by the software testing research community in relation to the adoption of LLMs. Given the huge explosion of interest in this topic, we expect that the literature is not yet mature enough to highlight effective and long-term conclusions clearly, nor is it stable enough to allow for a comprehensive review. Nevertheless, we believe that among such an impressive production of ideas and proposals, it can be valuable to start outlining those results that appear the most promising, reporting their remarked challenges, and then forecasting their expected impact on the future of software testing research. 
Thus, we came to the idea of a literature study performed as a semi-systematic literature review (SSLR)~\cite{Snyder2019}. The outcomes are presented in a roadmap for researchers interested in LLM-based testing.

An SSLR provides an overview of a broad research area, in which, quoting Snyder~\cite{Snyder2019}, reviewing ``\textit{every single article that could be relevant to the topic is simply not possible}''. Thus, the strict requirements behind a more rigorous systematic literature review, as presented in~\cite{kitchenham2004procedures}, can be somewhat relaxed.  Nevertheless, a transparent process must be followed, so that readers have the information for judging how the arguments and conclusions have been reached, ``\textit{from a methodological perspective}''~\cite{Snyder2019}. Thus, in the following, we explain the process followed in our SSLR.

Specifically, based on our four objectives SO1 -- SO4, as stated in the Introduction, our review of the literature has been guided by the following research questions (in parentheses, a forward reference to the sections in which they are addressed):
 
\begin{description}
\item[\textbf{RQ1:}] What are the main research categories in LLM-based testing? (see Section~\ref{sec:corpora})
\item[\textbf{RQ2:}] Which results have been achieved so far in those categories? (see Section~\ref{sec:llmUsage})
\item[\textbf{RQ3:}] What are the challenges, trends, and opportunities for research in LLM-based Testing? (see Section~\ref{sec:reflections:roadmap})
\item[\textbf{RQ4:}] How are LLMs impacting the software testing research landscape? (see Section~\ref{sec:reflections:dreams})
\end{description}

To answer such questions, we planned a search process across several digital libraries (DLs) and also performed a structured and bottom-up classification of the retrieved works, as in a systematic review. However, our SSLR relaxed two criteria usually followed in systematic studies. First,  we did not strongly pursue the idea of identifying a replicable \textit{closure-set} of all representative works on the investigated topic, acknowledging
that under these conditions, we might miss some relevant articles. Second, we relaxed the rigorousness of our investigation by including scientific works that are not yet peer-reviewed (e.g., published in arXiv). Nevertheless, we are confident that the semi-systematic process we defined can lead to 
an overview of the emerging ideas that have been so rapidly discussed during the last years, and to the first definition of a roadmap of LLM-based testing research, and more broadly on its impact on the future of software testing research.

The methodology we applied is structured in three main steps: \textit{planning} (see Section~\ref{sec:mapping:planning}), \textit{empirical collection} (see Section~\ref{sec:mapping:collection}), and \textit{incremental ad-hoc snowballing} (see Section~\ref{sec:mapping:snowballing}).

\subsection{Plan of the mapping}
\label{sec:mapping:planning}

The study focuses on articles written in English and that are indexed in at least one of the following DLs: ACM-DL (The ACM Guide to Computing Literature)\footnote{\url{https://dl.acm.org/search/advanced}} and IEEExplore\footnote{\url{https://ieeexplore.ieee.org/}}. Both DLs cover relevant scientific journals and conferences in software testing, and also index works in scientific journals/collections issued by publishers other than IEEE and ACM. In this sense, we expect to reach a significant (although not complete) portion of venues that could have published works in scope with the investigated topic.

Then we identified the macro-tokens to guide the querying of both DLs. Specifically, with the purpose of reaching a broad and relevant set, we made a generic search focusing on works just presenting (in their title) the combination of the token
``\texttt{test*}'' with either the token ``\texttt{llm*}'', or the token ``\texttt{large language model*}''.

For the analysis of the collected works, we devised a classification schema to highlight their main features. Specifically, the schema foresees to report (more details are provided in the next section): general information about the work (i.e., title, abstract, year of writing, type of publication, type of contribution), some feedback from the reviewer (i.e., own brief summary, a list on what is LLM-driven and what human needs to do, relevant research perspectives), information about the used technologies (i.e., the specific LLMs referred, any release of the tool implementing the proposed approach), information on how the approach has been validated (i.e., referred benchmarks, referred metrics on the observed outcomes), plus a space for any additional comment the reviewer wished to report. Finally, the schema includes two more fields (i.e., Category and LLM approach type) we used to tag each work based on a set of key dimensions. Specifically, the field ``Category'' aims to describe the main objective of the work with respect to software testing activities; starting from an initial set of alternatives inspired by the review by Wang et al.~\cite{WangTSE2024}, we planned to follow a bottom-up observation of evidence from the collected works, and elaborate iteratively coherent categories of testing objectives until reaching an agreed complete set. Instead, the ``LLM approach type'' aims to model the kind of interaction foreseen against the LLM: we had initially considered to distinguish between approaches that interact with a pre-trained LLMs just by casting proper prompts, and approaches that instead perform an ad-hoc fine-tuning. After reading some works, it became clear that these two categories were not sufficient to classify all works, and we added two more ``hybrid'' types of interactions, as explained in Section~\ref{sec:corpora}.

\subsection{Empirical collection and analysis}
\label{sec:mapping:collection}

On late March 2025, we queried the identified DLs by using the macro-tokens reported in Section~\ref{sec:mapping:planning}. These results represented the initial pool of articles we used in order to build the corpora for elaborating our roadmap. As reported in Section~\ref{sec:mapping:snowballing}, successively, we leveraged an incremental snowballing process that expanded our exploration also to other works.
 
At least two authors reviewed each retrieved article.  
In a first iteration, the articles were processed by only analyzing the title and abstract. Such a preliminary analysis aimed to discard works that were clearly out of scope for our study: for example, accidentally matching the macro-tokens in their title, or addressing undesired topics (e.g., hardware testing). All the authors have collectively made the final decision about those articles marked as discarded. In a second iteration, the reviewers processed the remaining works by considering their whole contents; all the information from the classification schema has been revised and detailed. In addition, reviewers proposed an initial bottom-up tagging of the articles with respect to the research dimensions described above. Periodically, all authors met in order to revise together the current set of adopted tags, to refine/align their definitions, and to generalize/specialize emerging classifications and LLM approach types. During these meetings, the authors also discussed further doubtful articles, some of which, after a deeper analysis, were discarded.

\subsection{Snowballing}
\label{sec:mapping:snowballing}

The last step of the applied methodology aimed to include relevant works that escaped from the querying of the DLs. We structured an incremental \textit{ad-hoc} snowballing procedure which covered any kind of scientific publication, also not peer-reviewed articles from arXiv: we took this decision in view of the rapid growth of the topic, and in the interest of having in our roadmap an up-to-date vision of ongoing research. Specifically, for each work currently included in the corpora that has not been previously processed, we applied both \textit{backward} and \textit{forward} snowballing. In the former case, we processed all its references. In the latter, we looked for all the works that GScholar reports in the collection ``\textit{Cited by}''~\footnote{performed between March-May 2025 on:~\url{https://scholar.google.com/}}. For both cases, we considered only those works whose title suggests some relevance to the goal of the research roadmap.

All articles resulting from each snowballing iteration were considered new articles that could potentially be included in the corpora of this study. Thus, they have been classified by at least two reviewers according to the common classification schema described in Section~\ref{sec:mapping:planning}. In addition, the authors met periodically to discuss works that were proposed for discarding and to revise the current status of the tags based on the bottom-up classification procedure. Thus, during snowballing, the authors iteratively analyzed the current clustering of articles in potential categories, and they collaboratively processed the definitions of each identified tag (e.g., by refining or proposing new ones). 

We classified the adopted snowballing procedure as \textit{ad-hoc} due to the implemented stopping condition. Indeed, we considered the snowballing complete (for the purpose of our roadmap) either when all the articles in the corpora have been processed, or when we observed that the set of tags referred to in the field ``\textit{Category}'' converged. 
Specifically, after a couple of iterations, we observed that the new works found through snowballing did not lead to any further change in the organization of the dimension ``\textit{Category}'', and thus we deemed that our corpora could sufficiently represent the main topics undergoing investigation in the current literature.

%% file: 04_papercomporae.tex
\section{Article Corpora}
\label{sec:corpora}

\begin{figure}[hbt]
 \centering
 \includegraphics[width=0.9\textwidth]{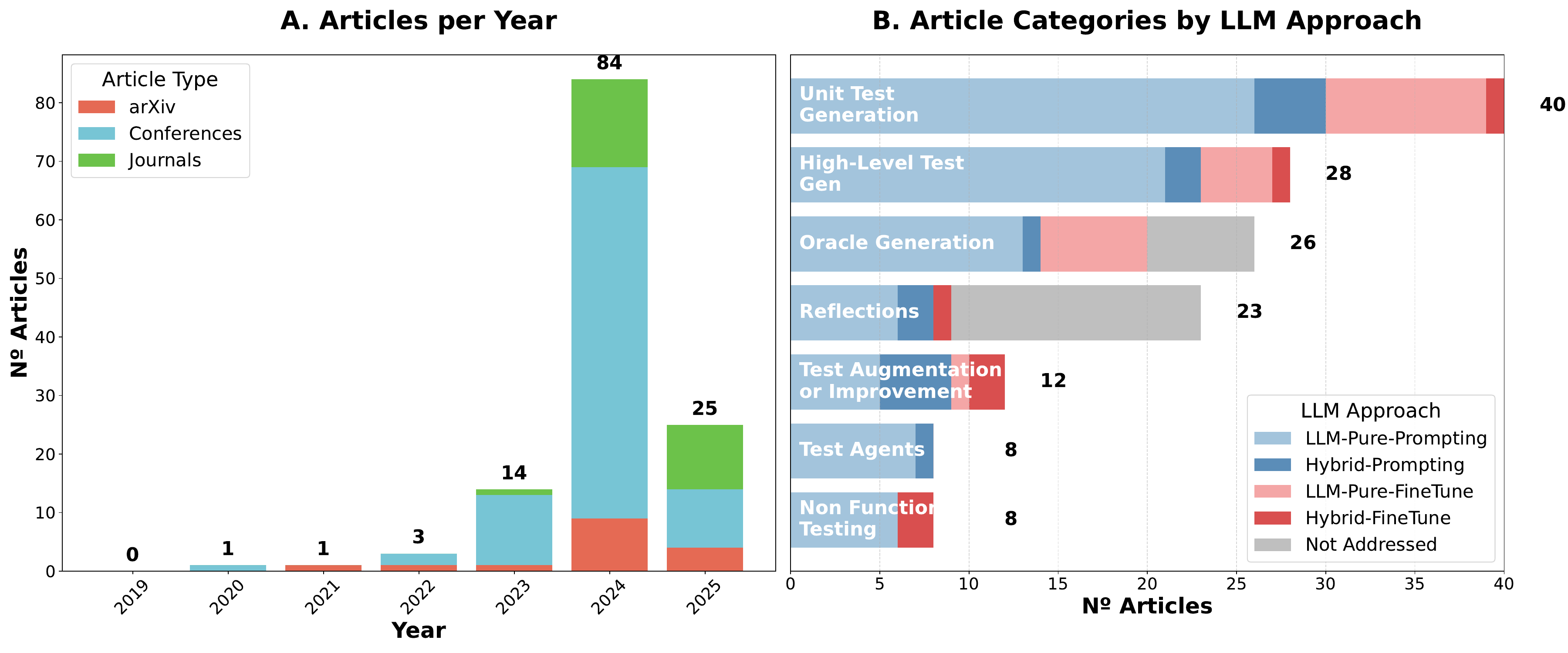}
 \caption{Publication years, sources, categories, and type of LLM approach}
 \label{fig:yearsandcat}
\Description[Publication years, sources, categories, and type of LLM approach]{Publication years, sources, categories, and type of LLM approach}
\end{figure}

In this section, we provide a classification of the \numarticles articles collected through the above process based on the year of publication or of their inclusion on arXiv, and the type of contribution (Journal, Conference, or arXiv). Fig.~\ref{fig:yearsandcat} illustrates the distribution of publications from 2020 to 2025, highlighting the proportions of contributions from Journals (green), Conferences (blue), and arXiv (red). No publications were found for 2019, and four distinct works were identified from 2020 to 2022.
In 2023, the number of retrieved publications increased by nearly 470\% (14), five times more in 2024 (84), and 25 articles were collected in 2025 up to the collection date. 

The type of contributions varies significantly. In the initial years, most publications appeared in conferences, particularly core venues such as ICSE, FSE, AST, and ASE, as well as on arXiv. However, since 2023, there has been a notable increase in articles about LLM-based testing published in journals, including high-ranking journals like TSE, IST, ACM TOSEM, and JSS.
Moreover, while writing this article, we observed a clear trend indicating that many publications we had initially found in arXiv were presented at various conferences and journals. This finding highlights the importance of regularly updating our article corpora to maintain the accuracy and relevance of our research materials.

As anticipated in Section~\ref{sec:mapping:planning}, we have classified our articles based on different categories, in relation to the LLM usage from a testing perspective. To establish these categories, we initially drew inspiration from the classification proposed by Wang et al.~\cite{WangTSE2024}, who identified the following categories: unit test generation, test oracle generation, system test input generation, bug analysis, debug, and repair. Of these, we considered:
\begin{itemize}
    \item \textbf{Unit Test Generation:} referring to articles leveraging LLMs for the generation of test cases (or related activities) at the unit test level.
    \item \textbf{High-Level Test Generation:} as above, this group includes works focusing on test cases generation, but at higher levels, e.g., system testing, end-to-end testing, etc.
    \item \textbf{Oracle Generation:} this group of articles addresses the derivation of oracles, such as, among others, the generation of test assertions and metamorphic testing. 
\end{itemize}

Instead, we did not consider the categories of bug analysis, debugging, and program repair, as the scope of our study covers the very stage of testing in search of potential failures, but not the following stages aimed at locating or repairing possible bugs. During our literature review and analysis, further categories emerged as we found further different uses of LLMs: we individually annotated new categories. We then collectively decided on their grouping and labelling in plenary periodic meetings. In summary, we introduced four new categories as follows:
\begin{itemize}
    \item \textbf{Test Augmentation or Improvement:}  this group includes methods that use LLMs to improve traditional test approaches, e.g., by enhancing code coverage or failure detection effectiveness.
    \item \textbf{Non-Functional Testing}: articles that discuss software testing of attributes not related to functionality (e.g., usability, efficiency, security) using LLMs.
    \item \textbf{Test Agents:} covers contributions that present one or more software agents that serve as intermediaries between testers and the testing process, and leverage LLMs for their operation.
\end{itemize}

In addition, we also introduced another category for articles covering LLM-based testing but not proposing novel approaches, e.g., 
\begin{itemize}
     \item \textbf{Reflections:} covers contributions that review the literature or provide an empirical or opinionated analysis of the existing approaches in LLM-based testing.
\end{itemize}

As introduced in Section~\ref{sec:mapping:planning}, in addition to the category, we differentiate between four distinct types of LLM approaches based on the synergies with existing tools and the models used. These approaches include:

\begin{itemize}
    \item \textbf{LLM-Pure prompting:} the proposed approach or tool uses a general-purpose LLM without fine-tuning it.
    \item \textbf{Hybrid Prompting:} the general-purpose LLM without fine-tuning is combined with a state-of-the-art (henceforth referred to as SOTA) tool to improve its performance.
    \item \textbf{LLM-Pure Fine-tune:} the proposed approach or tool uses a fine-tuned LLM for a specific purpose or context.
    \item \textbf{Hybrid Fine-tune:} the proposal combines a SOTA tool with a fine-tuned LLM for the specific purpose.
\end{itemize}

 The right part of Fig.~\ref{fig:yearsandcat} represents the article categories with the different LLM approaches employed in each category. All categories have publications on the four types of LLM approach, with a predominance of the pure-prompting approaches over fine-tuning and hybrid approaches.

 Table \ref{tab:article_summary} presents the different articles considered in the roadmap grouped by their category and ordered by the number of contributions. Each article has a unique identifier (PXX) for identification in the supplementary material. \footnote{Replication package and supplementary material publicly available: https://github.com/giis-uniovi/llm-testing-roadmap-rp }.  The total articles in the right column is higher than those retrieved, as some belong to multiple categories (e.g., high-level test generation with test agents and reflections with test oracles).

\input{sheet2_1}

\begin{figure}[hbt]
\centering
\includegraphics[width=0.9\textwidth]{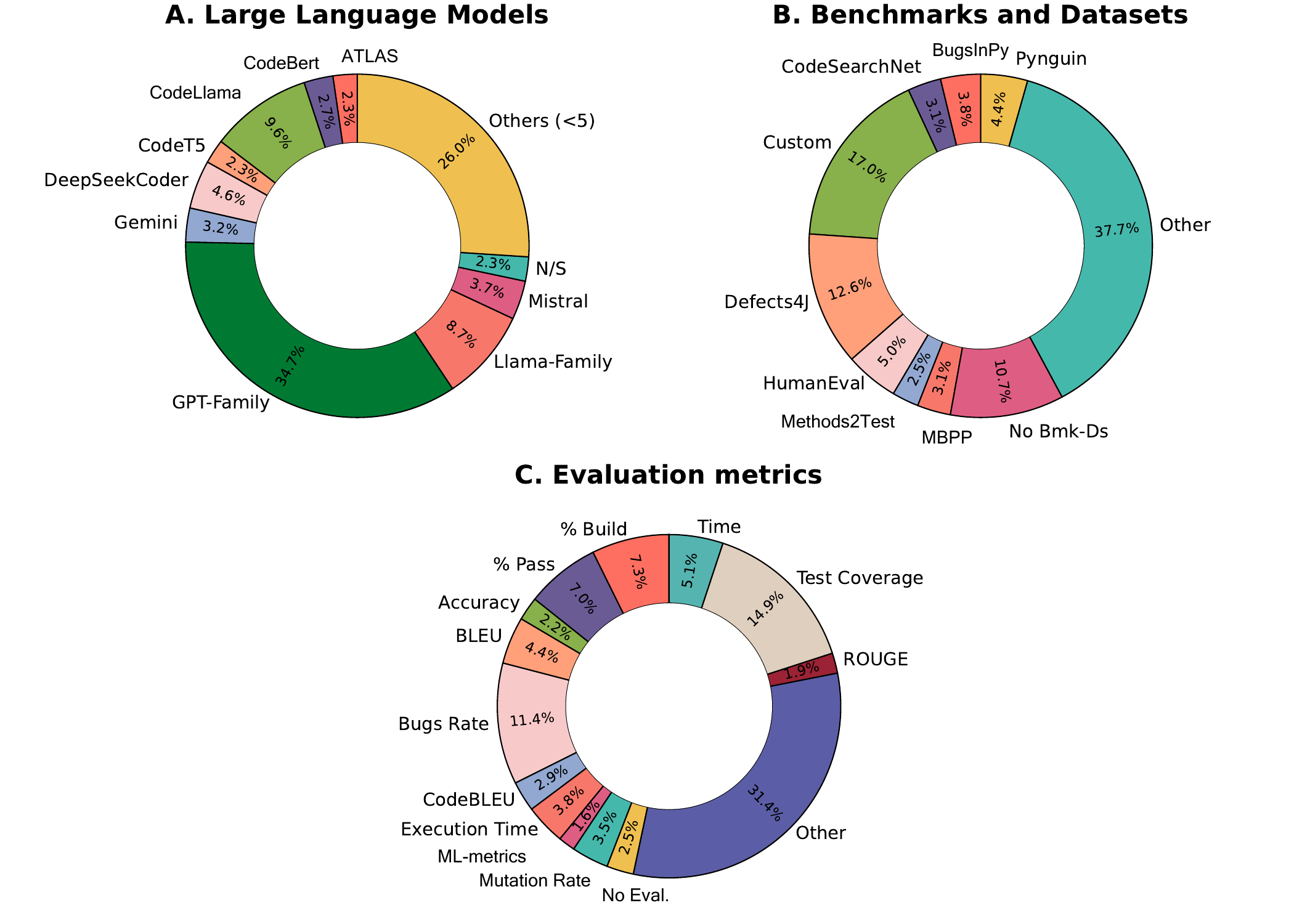}
\caption{Usage of LLM models, benchmarks and datasets} \label{fig:llmmodelbenchdataset}
\Description[Usage of LLM models, benchmarks and datasets]{Usage of LLM models, benchmarks and datasets}
\end{figure}

In each article where an approach is evaluated or presented, we have identified the models used, the benchmarks and datasets utilized in the evaluation, as well as the metrics adopted. The pie charts in Fig.~\ref{fig:llmmodelbenchdataset} depict the different LLMs, benchmarks, and metrics used.

In particular, among the models used in the articles (Fig.~\ref{fig:llmmodelbenchdataset},~\textcircled{A}), the OpenAI GPT family in its different versions (3.5-turbo, 3.5, 4, 4o, 4o-mini) is quite preferred, appearing in over half (35\%) of the articles considered. Besides, the open-source Llama family (Llama, Llama3, and Code Llama) enjoys significant popularity, appearing in nearly 18.3\% of the articles, with DeepSeek, Mistral, and Gemini contributing with another 4.6\%, 3.7\%, and 3.2 \% respectively. To improve the diagram's visibility, we have grouped the remaining models,  used in fewer than four articles, into the Others category, including models like Starcoder-2, BART, or QWen.

The articles employ a broad range of benchmarks (Fig.~\ref{fig:llmmodelbenchdataset},~\textcircled{B}) depending on the programming language and the category considered. In the Python language, benchmarks like Pynguin, BugsinPy, and MBPP are popular, appearing in 10.7\% of the articles. In Java, the Defects4J benchmark is present in 12.6\% of the articles, while HumanEval, a benchmark released by OpenAI to measure coding skills for their models, appears in 5\% of the articles. Several articles (17\%) create their custom datasets, whereas the remaining ones utilize benchmarks like GitBug, CodeSearchNet, and APPs.

Finally, regarding the metrics (Fig.~\ref{fig:llmmodelbenchdataset},~\textcircled{C}), they are dependent on the output generated by the LLMs. When the output is in natural language, such as input data or test scenarios, some studies utilize NLP metrics like ROUGE or BLEU, along with the time taken by the model. In contrast, when the output is code, such as test cases, researchers typically adopt SOTA metrics. These may include the number of bugs found, mutation rates, various types of coverage, and the number of test cases that pass or build successfully. Additionally, they may use traditional machine learning metrics like F1 score, accuracy, or precision, each providing different insights into true positives (TP), false positives (FP), true negatives (TN), and false negatives (FN). Finally, many works also employ other metrics, some of which are specifically designed for the approaches being evaluated.

%% file: sheet2_1.tex
\begin{table}[hbt]
    \caption{Distribution of Articles by Categories}
    \label{tab:article_summary}
    \centering
    \begin{adjustbox}{max width=\linewidth}
    \begin{tabular}{m{8.355em} m{26.355em} >{\centering\arraybackslash}m{3.355em} }
         \hline
        \textbf{Category} & \textbf{Articles} & \textbf{TOTAL} \\ \hline
        Unit Test Generation & 
        \textbf{P03}~\cite{Schafer2024}, \textbf{P07}~\cite{Plein2024}, \textbf{P09}~\cite{Vikram2024}, \textbf{P10}~\cite{ZhangICSE2025}, \textbf{P11}~\cite{ChenFSE2024}, \textbf{P13}~\cite{Ryan2024}, \textbf{P20}~\cite{Yuan2024}, \textbf{P24}~\cite{WangTESTEVAL2024}, \textbf{P25}~\cite{Jiang2024}, \textbf{P28}~\cite{Bhatia2024}, \textbf{P32}~\cite{Huang2024}, \textbf{P34}~\cite{OuedraogoASE2024}, \textbf{P37}~\cite{Liu2025}, \textbf{P40}~\cite{Yin2025}, \textbf{P43}~\cite{Lops2024}, \textbf{P46}~\cite{Hoffman2024}, \textbf{P55}~\cite{Taherkhani2024}, \textbf{P56}~\cite{Chen2023}, \textbf{P57}~\cite{Gao2025}, \textbf{P60}~\cite{ChenArxiv2024}, \textbf{P66}~\cite{Lin2024}, \textbf{P74}~\cite{Tufano2021}, \textbf{P76}~\cite{Nie2023}, \textbf{P78}~\cite{Rao2024}, \textbf{P81}~\cite{Alagarsamy2024}, \textbf{P83}~\cite{Pan2025}, \textbf{P87}~\cite{Ni2024}, \textbf{P88}~\cite{Zhong2024}, \textbf{P98}~\cite{Garlapati2024}, \textbf{P102}~\cite{Wanigasekara2024}, \textbf{P108}~\cite{Konuk2024}, \textbf{P112}~\cite{Wang2024}, \textbf{P117}~\cite{Kang2023}, \textbf{P119}~\cite{Jiri2024}, \textbf{P124}~\cite{PaduraruJournal2024}, \textbf{P127}~\cite{Yang2025}, \textbf{P128}~\cite{Munley2024}, \textbf{P129}~\cite{XiaoInt2024}, \textbf{P138}~\cite{Yeh2024}, \textbf{P142}~\cite{Paduraru2024} & 40 \\ \hline
        High-Level Test Generation & \textbf{P15}~\cite{Chetan2024}, \textbf{P18}~\cite{XueISSTA2024}, \textbf{P19}~\cite{Karmarkar2024}, \textbf{P30}~\cite{Augusto2025}, \textbf{P33}~\cite{Liu2024}, \textbf{P48}~\cite{Bergsmann2024}, \textbf{P58}~\cite{Yu2023}, \textbf{P65}~\cite{Liu2023}, \textbf{P68}~\cite{Alian2025}, \textbf{P70}~\cite{SuYanqi2024}, \textbf{P71}~\cite{LiuGUI2024}, \textbf{P99}~\cite{Kang2024}, \textbf{P100}~\cite{Shakthi2024}, \textbf{P103}~\cite{Karpurapu2024}, \textbf{P105}~\cite{Zhang2025}, \textbf{P106}~\cite{Feng2024}, \textbf{P107}~\cite{Su2024},\textbf{P113}~\cite{SantosSBQS2024}, \textbf{P114}~\cite{Yoon2024}, \textbf{P115}~\cite{Le2024}, \textbf{P121}~\cite{WangICSME2024}, \textbf{P122}~\cite{Yin2024}, \textbf{P125}~\cite{Petrovic2024}, \textbf{P136}~\cite{SuQingram2024}, \textbf{P137}~\cite{XuQRS2024}, \textbf{P138}~\cite{Yeh2024}, \textbf{P139}~\cite{LiDTPI2024}, \textbf{P140}~\cite{Almutawa2024} & 28 \\ \hline
        Oracle Generation & \textbf{P23}~\cite{Molina2024}, \textbf{P35}~\cite{Khandaker2025}, \textbf{P49}~\cite{Tufano2022}, \textbf{P50}~\cite{Watson2020}, \textbf{P51}~\cite{Primbs2025}, \textbf{P62}~\cite{Konstantinou2024}, \textbf{P63}~\cite{Hayet2025}, \textbf{P64}~\cite{Hossain2025}, \textbf{P67}~\cite{Noor2023}, \textbf{P75}~\cite{Elizabeth2022}, \textbf{P76}~\cite{Nie2023}, \textbf{P77}~\cite{Soneya2023}, \textbf{P79}~\cite{Jiho2024}, \textbf{P80}~\cite{Zhongxin2023}, \textbf{P81}~\cite{Alagarsamy2024}, \textbf{P82}~\cite{Yibo2024}, \textbf{P84}~\cite{ZhangTSE2025}, \textbf{P85}~\cite{Hailong2024}, \textbf{P87}~\cite{Ni2024}, \textbf{P89}~\cite{Hung2023}, \textbf{P90}~\cite{Yifan2023}, \textbf{P91}~\cite{Seung2024}, \textbf{P92}~\cite{Congying2024}, \textbf{P93}~\cite{Madeline2024},  \textbf{P144}~\cite{Huynh2024}, \textbf{P145}~\cite{TsigkanosICCS2024} & 26 \\ \hline
        Reflections & \textbf{P01}~\cite{Fei2024}, \textbf{P02}~\cite{WangTSE2024}, \textbf{P04}~\cite{Santos2024}, \textbf{P08}~\cite{Savage2024}, \textbf{P14}~\cite{Mathews2024}, \textbf{P16}~\cite{OuedraogoArxiv2024}, \textbf{P22}~\cite{Braberman2024}, \textbf{P23}~\cite{Molina2024}, \textbf{P31}~\cite{Russo2024}, \textbf{P36}~\cite{Boukhlif2024}, \textbf{P41}~\cite{Li2025}, \textbf{P42}~\cite{Shang2024}, \textbf{P44}~\cite{Abdullin2025}, \textbf{P45}~\cite{Siddiq2024}, \textbf{P59}~\cite{Fraser2025}, \textbf{P72}~\cite{Zheng2025}, \textbf{P77}~\cite{Soneya2023}, \textbf{P79}~\cite{Jiho2024}, \textbf{P80}~\cite{Zhongxin2023}, \textbf{P82}~\cite{Yibo2024}, \textbf{P84}~\cite{ZhangTSE2025}, \textbf{P130}~\cite{Shi2024}, \textbf{P143}~\cite{He2024} & 23 \\ \hline
        Test Augmentation or Improvement & \textbf{P05}~\cite{Alshahwan2024}, \textbf{P12}~\cite{Lemieux2024}, \textbf{P17}~\cite{Myeongsoo2024}, \textbf{P21}~\cite{WangArxiv2024}, \textbf{P38}~\cite{Altmayer2025}, \textbf{P39}~\cite{Dakhel2024}, \textbf{P47}~\cite{Fatima2024}, \textbf{P54}~\cite{ZhangAST2025}, \textbf{P61}~\cite{Meng2024}, \textbf{P110}~\cite{LiuISSRE2024}, \textbf{P135}~\cite{Rahman2024}, \textbf{P142}~\cite{Paduraru2024} & 12 \\ \hline
        Test Agents & \textbf{P26}~\cite{Feldt2023}, \textbf{P29}~\cite{Bouzenia2024}, \textbf{P98}~\cite{Garlapati2024}, \textbf{P114}~\cite{Yoon2024}, \textbf{P116}~\cite{Jorgensen2024}, \textbf{P131}~\cite{Bianou2024}, \textbf{P133}~\cite{Wu2024}, \textbf{P140}~\cite{Almutawa2024} & 8 \\ \hline
        Non Functional Testing & \textbf{P111}~\cite{Happe2023}, \textbf{P120}~\cite{Calvano2025}, \textbf{P126}~\cite{Pasca2025}, \textbf{P131}~\cite{Bianou2024}, \textbf{P132}~\cite{Deng2024}, \textbf{P133}~\cite{Wu2024}, \textbf{P136}~\cite{SuQingram2024}, \textbf{P146}~\cite{Chen2024} & 8 \\ \hline
    \end{tabular}
    \end{adjustbox}
\end{table}

%% file: 05_llmUsage.tex
\section{Results in LLM-based testing research}
\label{sec:llmUsage}

In this section, we give a brief summary of the status of research in LLM-based testing, covering from Section~\ref{sec:llmusage:unittestgen} to Section~\ref{sec:llmusage:nofunctest}, respectively, the six categories\footnote{The 7th category of Reflections is covered in Section~\ref{sec:reflections}}:  Unit Test Generation, High-level Test Generation, Oracle Generation, Test Augmentation or Improvement, Non-Functional Testing, and Test Agents. 

For each category, we include a scheme that attempts a generic process of how LLMs are used in works belonging to the category. This scheme itself is provided as a contribution to grasp a quick overview of the topic.

\subsection{Unit Test Generation}
\label{sec:llmusage:unittestgen}
\begin{figure}[htb]
\centering
\includegraphics[width=0.9\textwidth,trim=0 120 0 100, clip]{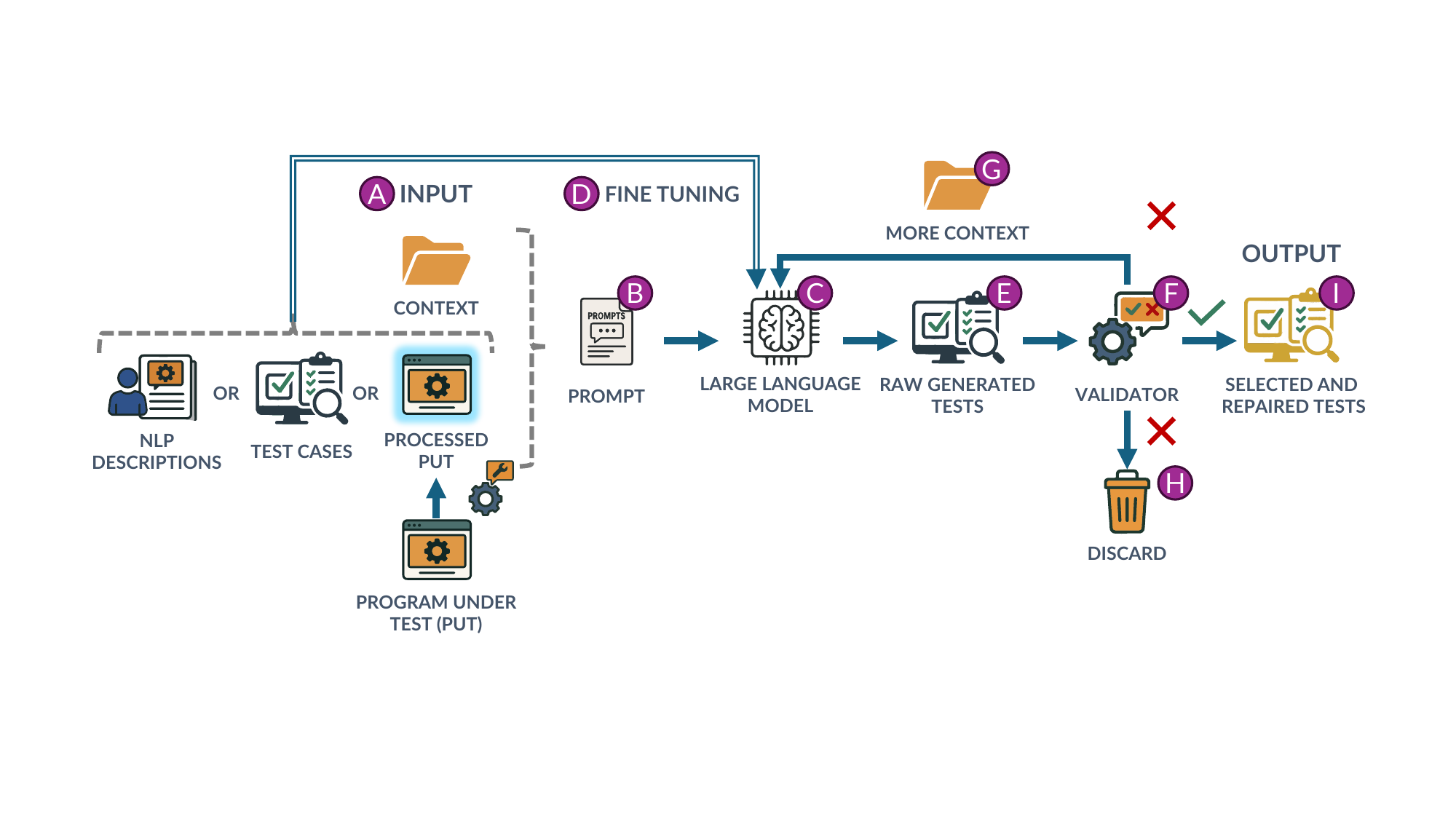}
\caption{Unit Test Generation Process} 
\label{fig:unitgenprocess}
\Description[Unit Test Generation Process]{Unit Test Generation Process}
\end{figure}

Since the beginning, LLMs have largely been used for the automated generation of unit tests. 
Table \ref{tab:article_summary} classifies \numunit articles in LLM-based unit test generation. Fig.~\ref{fig:unitgenprocess} depicts the process for unit test case generation.

 The input for LLM-based unit test generation (Fig.~\ref{fig:unitgenprocess}~\textcircled{A}) can belong to three types (differently combined): (1) the program under test (PUT) (most commonly)~\cite{Schafer2024, ZhangICSE2025, Ryan2024, Bhatia2024, Huang2024, OuedraogoASE2024, Yin2025, Lops2024, Hoffman2024, Lin2024, Tufano2021, Nie2023, Rao2024, Alagarsamy2024, Pan2025, Ni2024, Garlapati2024, Konuk2024, Wang2024, Jiri2024, PaduraruJournal2024, Yang2025, XiaoInt2024, Paduraru2024}, (2) functionality textual descriptions like bug reports~\cite{Plein2024, Kang2023}, documentation ~\cite{Vikram2024, Yin2025, Gao2025}, commit messages, bug patches ~\cite{Liu2025} or requests in natural language~\cite{Munley2024}, or (3) pre-existing test cases~\cite{ChenArxiv2024, Zhong2024, Yeh2024}.
Some works integrate such inputs with more context to enhance the unit test generation. Examples of additional information provided into the context are the documentation~\cite{Schafer2024, Yang2025}, examples of the desired output~\cite{Schafer2024}, additional source (e.g. the test method signature, the code of methods invoked inside the PUT)~\cite{ChenFSE2024, Gao2025, Tufano2021, Pan2025, XiaoInt2024}, error feedback~\cite{Schafer2024} or examples of test cases~\cite{Kang2023}. 

This context is placed into the crafted prompt (Fig.~\ref{fig:unitgenprocess}~\textcircled{B}), which is used to query the LLM to generate the unit test cases. The LLM (Fig.~\ref{fig:unitgenprocess}~\textcircled{C}) can be a general-purpose model (e.g., ChatGPT, CodeLlama, Gemini) or a fine-tuned version (Fig.~\ref{fig:unitgenprocess}~\textcircled{D}). The fine-tuning in unit test case generation is done with examples of test cases~\cite{ZhangICSE2025}, PUTs~\cite{Tufano2021}, pairs of PUTs and test cases~\cite{Hoffman2024, Nie2023, Rao2024, Alagarsamy2024, Paduraru2024, Munley2024, PaduraruJournal2024}.

The LLM produces one or more test cases (Fig.~\ref{fig:unitgenprocess}~\textcircled{E}). Some approaches include a validation phase (Fig.~\ref{fig:unitgenprocess}~\textcircled{F}), to either (1) discard invalid test cases (Fig.~\ref{fig:unitgenprocess}~\textcircled{H}), or  (2)  iteratively fix or improve the test case generation using the LLM-in-the-loop. In unitary test case generation, additional context generated in the validation phase is included in the LLM (Fig.~\ref{fig:unitgenprocess}~\textcircled{G}) to guide and improve the generated test cases. Examples of additional information that is included in the context is the test coverage~\cite{Jiang2024, Ryan2024, Bhatia2024, Gao2025, ChenArxiv2024, XiaoInt2024}, the test syntax errors, test smells, failures traces, compilation problems or bug detection rate~\cite{Schafer2024, ChenFSE2024, Ryan2024, Yuan2024, Lops2024, Lin2024, Alagarsamy2024, Pan2025, Garlapati2024, PaduraruJournal2024, Yang2025, Kang2023}. 
The final output from one to several validated unit test cases  (Fig.~\ref{fig:unitgenprocess}~\textcircled{I}).

A clear trend in unit test generation is the integration of the LLM with existing tools for its enhancement. For instance some authors have combined the LLMs data generation tools~\cite{Jiang2024}, enhanced LLMs with ML approaches to detect the hallucinations and improve the LLM output~\cite{Taherkhani2024}, complement the existent approaches, e.g., enabling to reach situations that the SOTA tools are not able to cover~\cite{ChenArxiv2024}, combine the LLM with the existant mutation testing tools~\cite{Zhong2024} or complement enhance algorithms~\cite{Yang2025} that those tools use. 

The different approaches and tools employ a broad range of LLMs: there's a predominance of the GPT-Family, but also open source models like the Llama-family, code purpose LLMs like CodeLlama and StarCoder, and also introduce new models tailored for specific unit test generation tasks like TECO~\cite{Nie2023}, CAT-LLM~\cite{Rao2024}, or  AthenaTest~\cite{Tufano2021}. The performance of these models is evaluated with different metrics, as unitary test generation, they have access to the PUT code, and they can employ metrics like test coverage, mutation rate, bug detection rate, and number of test smells. However, they also use LLM-crafted metrics like NLP metrics as BLEU, ROUGE, and XMatch, or execution metrics, like the percentage of test cases that run, build, and pass correctly. 
 
 In general, articles in test unit generation use the datasets and benchmarks employed in the SOTA approaches, depending on the programming language in which the test cases are generated. Examples of those benchmarks are Defects4J Pynguin, BugsInPy, Codesearchnet, Apps, SF11, or MBPP. Also, several benchmarks have been published for the specific purpose of LLM-based test unit generation, like TESTEVAL~\cite{WangTESTEVAL2024}, human-crafted Python programs to evaluate performance in unit gen, Classes2Test~\cite{Lops2024} derived from Methods2Test with the entire class instead of solely the test method, or Testsuite V\&V~\cite{Munley2024}

Finally, some authors have presented tools like CasModaTest~\cite{Ni2024}, JSONTestGen~\cite{Zhong2024}, A3Test~\cite{Alagarsamy2024}, CodeT~\cite{Chen2023}, AgoneTest~\cite{Lops2024}, ChatUnitTest~\cite{ChenFSE2024}, exLong~\cite{ZhangICSE2025}, or TestPilot~\cite{Schafer2024} which leverage the LLMs to generate unit tests or improve existing test suites.

\subsection{High-level Test Generation}
\label{sec:llmusage:highlevel}
\begin{figure}[htb]
\centering
\includegraphics[width=0.9\textwidth,trim=0 100 0 100, clip]{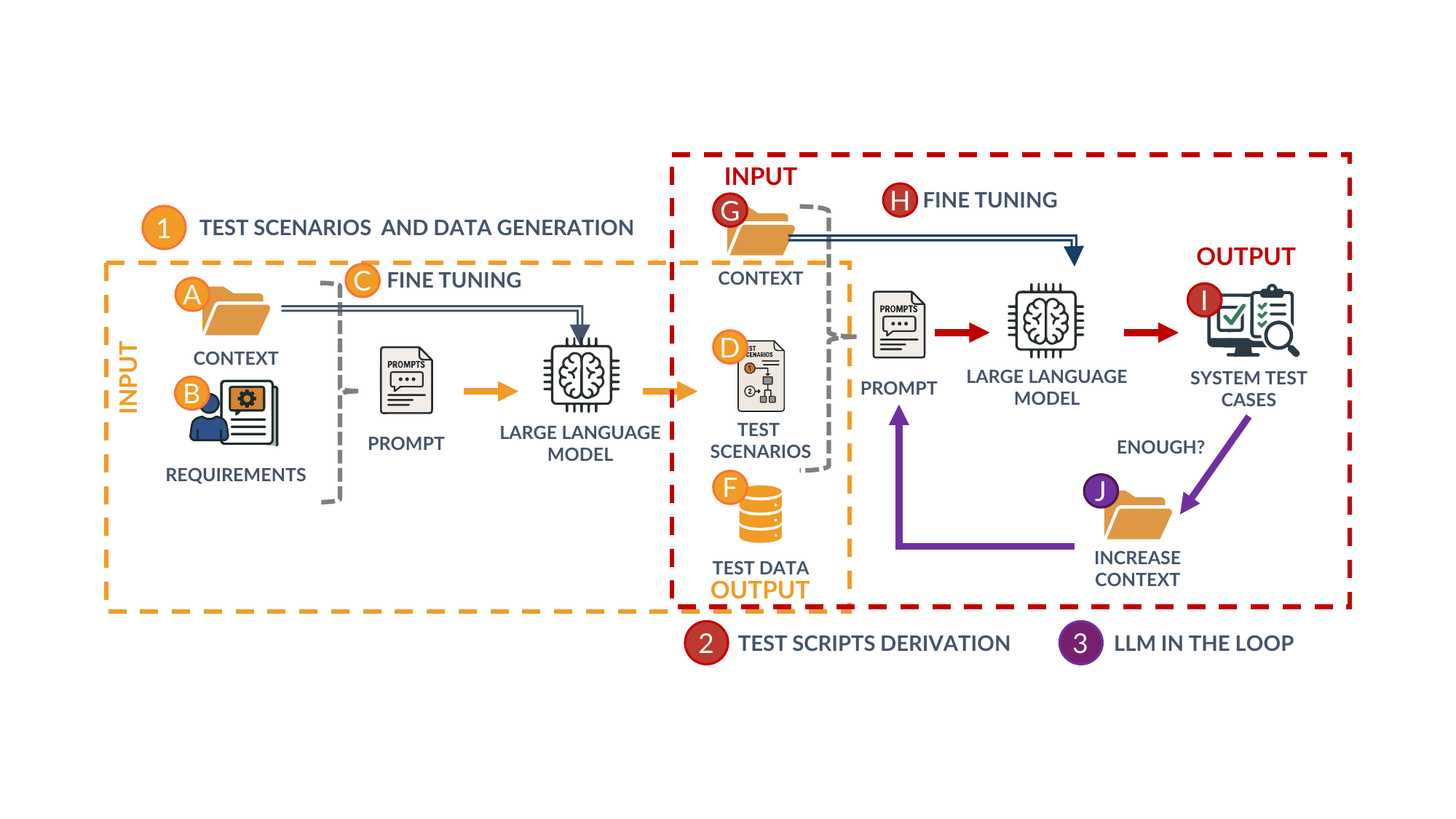}
\caption{High-Level LLM-based testing process} \label{fig:highendprocess}
\Description[High-Level LLM-based testing process]{High-Level LLM-based testing process}
\end{figure}

The High-level test generation is the second category with more contributions, ranging from \numhighend articles among the \numarticles reviewed.
High-level testing verifies the entire software system, from user interactions with the interface or calls to the API endpoint, to persistence operations like database insertions or deletions. The LLM-based High-End testing articles can be categorized into three groups: \textbf{Test Scenarios and Data Generation}, \textbf{Test Script Derivation}, and \textbf{Other usages}. Fig.~\ref{fig:highendprocess} represents the high-level test generation process, in yellow is the \textcircled{1} scenario and data generation, red is the  \textcircled{2} test script derivation, and in violet, we represent \textcircled{3} the use of LLM in the loop for the test script generation. Showcase that due to its heterogeneity, the works grouped in other usages~\cite{Yu2023, Yeh2024, Zhang2025, SuQingram2024} are not represented in the process model

\subsubsection{Test  Scenarios and Data Generation}

Comprehends the approaches and tools that aim to provide as output one or more test scenarios: \textit{``the situations or settings for a test item used as a basis for generating the test cases''}~\cite{ISO29119-1}, in natural or a human-machine readable language.
Represented in yellow color in Fig.~\ref{fig:highendprocess}, those approaches and tools use a pure-prompting LLM approach using prompting techniques to provide examples like Few-Shot or Chain of Thought. The requirements (Fig.~\ref{fig:highendprocess}~\textcircled{A}) examples provided are specifications, like the system requirements~\cite{Chetan2024, Petrovic2024, XueISSTA2024}, the user requirements~\cite{Augusto2025, Karpurapu2024}, but also less structured inputs like bug reports~\cite{Su2024}. 
The desired output is the set of actions to be performed (Fig.~\ref{fig:highendprocess}~\textcircled{D}), which can be expressed in natural language or in a structured language like Gherkin. The performance of the LLM generating the scenarios is measured in terms of the requirements covered~\cite{Augusto2025, XueISSTA2024}, traditional NLP metrics like METEOR, BLEU, or ROUGE~\cite{Chetan2024} when are in natural language, the syntax errors~\cite{Karpurapu2024}, and the cost in terms of money or time~\cite{Petrovic2024}.

The \textbf{data generation} covers those articles strictly focused on generating test inputs or parameters to test the applications. The inputs used to generate test data are also requirements (Fig.~\ref{fig:highendprocess}~\textcircled{B})~\cite{Karmarkar2024, Shakthi2024}, but also valid inputs as context to be mutated-improved~\cite{Liu2024} or the context information (e.g., widget details, parents, leaf)~\cite{Liu2023}. These inputs are directly prompted to the LLM to retrieve the data (Fig.~\ref{fig:highendprocess}~\textcircled{E}), but some works also use them in desired input-output pairs to fine-tune the model (Fig.~\ref{fig:highendprocess}~\textcircled{C}) and improve the performance. The quality of those generated inputs is measured in terms of test correctness, test passing, and the number of bugs detected.

There are no SOTA benchmarks in the test data and scenarios generation, and the LLMs used belong to the GPT and Llama families, as well as less popular LLMs like ChatGLM, FinBert, or  Palm2.

\subsubsection{Test Script Derivation}

The test script derivation in LLM-based High-to-end testing is more heterogeneous; we devise three different groups: \textit{test generation}, \textit{test generation with the LLM in the loop}, and \textit{data generation} approaches. The test script derivation is depicted in the right part of the Fig.~\ref{fig:highendprocess}, divided into the test script derivation (Fig.~\ref{fig:highendprocess}~\textcircled{2}) in red color and the LLM in the loop (Fig.~\ref{fig:highendprocess}~\textcircled{3}) in violet color

The \textbf{test generation} approaches get as input how the application should behave(Fig.~\ref{fig:highendprocess}~\textcircled{D}), which can be specified in a natural language scenario~\cite{Augusto2025, Bergsmann2024, SantosSBQS2024}, documentation/specifications~\cite{Le2024, XuQRS2024}, or usage patterns~\cite{Feng2024}, which are prompted to the LLM with examples of the desired output (Fig.~\ref{fig:highendprocess}~\textcircled{G}). 
Other works fine-tune the model (Fig.~\ref{fig:highendprocess}~\textcircled{H}) with examples of test cases with the domain knowledge~\cite{XueISSTA2024}, or examples of test cases in a programming~\cite{Kang2024} or natural language ~\cite{Yin2024}. The output of the script derivation process is the different test cases(Fig.~\ref{fig:highendprocess}~\textcircled{I}) in a programming language or in natural language (with the steps, inputs, and desired output).

The performance of those approaches is measured in different ways, black box approaches that doesn't have access to the application code employ how well covered the scenarios provided as input~\cite{XueISSTA2024}, the cost in terms of money-time~\cite{Feng2024, Bergsmann2024}, the human-effort required~\cite{Augusto2025}, NLP metrics~\cite{Kang2024} or the correctness of the test cases generated~\cite{Bergsmann2024}. The approaches that have the application code available mostly employ coverage in their different variations (e.g., line, method, branch)~\cite{XuQRS2024, Le2024, Yin2024}.

The \textbf{test generation with the LLM in the loop} (Fig.~\ref{fig:highendprocess}~\textcircled{3}) differs from the standard test generation approaches in the way that the LLM is used, employing a so-called LLM in the loop. This approach consists of iterating over the model several times, increasing the context (Fig.~\ref{fig:highendprocess}~\textcircled{J}) instead of a \textit{``single shot with examples''}, enhancing at each interaction the information available. The LLM in the loop approaches get as input GUI images or their context-description~\cite{LiuGUI2024, LiDTPI2024,  Liu2023}, the context of each interaction~\cite{WangICSME2024}, or the description of its functionality~\cite{Almutawa2024}. 

This information is prompted to a general-purpose LLM, but also some approaches fine-tune the LLM with desired inputs and answers~\cite{LiDTPI2024, Liu2023} (Fig.~\ref{fig:highendprocess}~\textcircled{H}). The output of these approaches (Fig.~\ref{fig:highendprocess}~\textcircled{I}) are a series of activities tasks to be performed, its performance is measured in terms of number of activities or states covered-explored~\cite{LiuGUI2024, Yoon2024, WangICSME2024}, bugs~\cite{LiuGUI2024}, the cost and number of lines generated~\cite{LiDTPI2024} and the number of tests that correct pass~\cite{Liu2023}.

In general, in the high-level \textbf{test generation}, no well-established benchmarks exist; some articles have proposed new benchmarks to evaluate the performance of LLM-based high-end test techniques~\cite{Alian2025}. Respecting the LLMs used, there is a predominance of the GPT-family (GPT 3.5, 4, 4o) but also open source models like the Llama family, Llava, Mistral, or Chat-GLM.

\subsubsection{Other Usages}
Other usages cover those articles that, by their nature, belong to high-level testing, but cannot be grouped in the previous two groups.  
In other usages, we have found articles that cover the test migration employing an LLM, migrating to a new framework or application. Between the articles considered, two works have explored the test migration capabilities: in~\cite{Yu2023} explores the inter-framework and inter-application LLM-migration capabilities, as well as the test generation providing input code, context, or GUI information. In~\cite{Yeh2024}, which has proposed the migration of integration-E2E test cases to contract test cases. The output of the model is those test cases migrated and whose performance is measured in terms of the test coverage or the number of dependencies-hallucinations.

Other works employ the LLMs to generate an initial set of test cases~\cite{Zhang2025} for then mutating them and discovering bugs, or fine-tuning the LLM to optimize the hyperparameters of another AI technique~\cite{SuQingram2024}. The performance of those approaches is evaluated in terms of the bug detection rate.
No well-established benchmarks exist in those categories, so custom ones are used in all contributions. The LLMs used belong to the GPT family, but also other less popular models like Qween2, Mistral, or Gemini.
 
\subsection{Oracle Generation}
\label{sec:llmusage:oraclegen}

One difficult part of software testing is the generation of oracles~\cite{barr2014}. Even when test generation tools such as Evosuite~\cite{Fraser2011} can detect most faults~\cite{Almasi2017}, they still rely on oracles~\cite{Fraser2010}. Users of these non-LLM tools do not always like the generated oracles and, in some cases, prefer to generate them manually~\cite{Almasi2017}. These non-LLM tools sometimes generate regression oracles~\cite{Xie2006} instead of taking advantage of the documentation. Note that the majority of GitHub projects have documentation in Natural Language~\cite{Pfeiffer2020}. LLMs can complement the traditional tools to generate test oracles that analyze documentation, among others.

\begin{figure}[htb]
\centering
\includegraphics[width=1\textwidth,trim=0 20 0 20, clip]{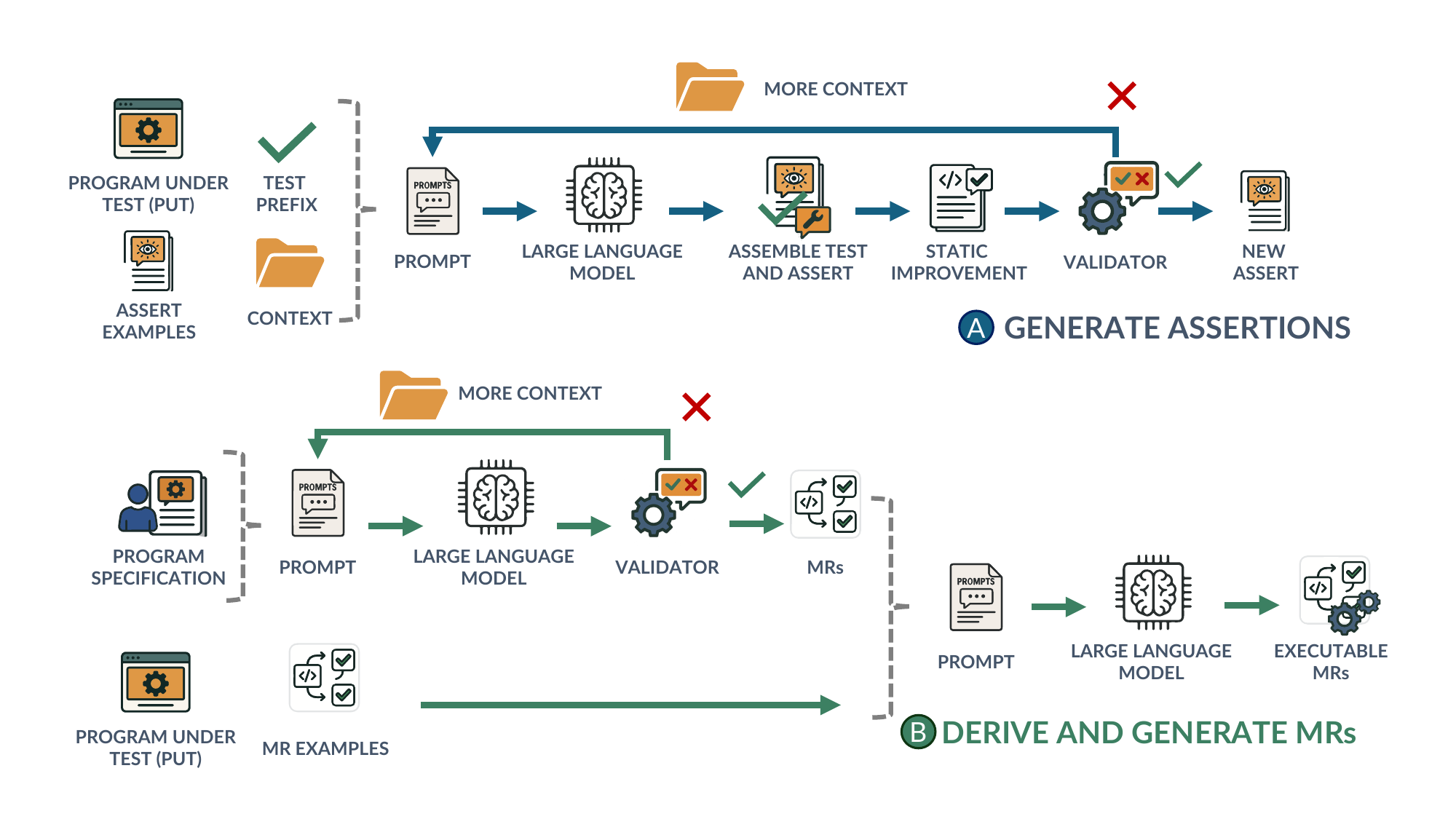}
\caption{LLM-based process for oracle generation through assertions~\textcircled{A} and MRs~\textcircled{B}} 
\label{fig:approaches_to_generate_oracles}
\Description[Assertions and MRs derivation LLM-based process]{Assertions and MRs derivation LLM-based process}
\end{figure}

Table~\ref{tab:types_oracles} classifies \numoracle articles that address the oracle problem using LLMs. The majority of contributions generate oracles through assert-like statements (58\text{\%}) and metamorphic relationships (19\text{\%}). The remaining articles, 23\text{\%}, are surveys or empirical studies on the current LLM-based approaches. Fig.~\ref{fig:approaches_to_generate_oracles} depicts how the 20 LLM-based approaches generate oracles in the form of assertions (Fig.~\ref{fig:approaches_to_generate_oracles}~\textcircled{A}) or metamorphic relations (Fig.~\ref{fig:approaches_to_generate_oracles}~\textcircled{B}).

\input{sheet3_1}

Table~\ref{tab:types_of_prompt_inputs_for_oracles} indicates the inputs used by LLM-based approaches to generate oracles. In general, the prompt is a combination of the program under test (75\text{\%}), the test prefix (60\text{\%}), some contextual information (55\text{\%}), and examples (10\text{\%}). The test prefix indicates to the LLM which is the goal of the test case. The prompt usually contains the program under test to help LLM understand what is being tested, sometimes with the code of the focal method (60\text{\%}), the signature (20\text{\%}), or providing other methods/classes (25\text{\%}). The LLMs also employ contextual information that is not directly in the code of the program, such as documentation (20\text{\%}), specification (30\text{\%}), or dependencies (10\text{\%}). The prompt can also contain examples of oracles (15\text{\%}) to guide the LLM in generating new ones. 

\input{sheet3_2}

According to Fig.~\ref{fig:approaches_to_generate_oracles}, the prompt is only one part of the whole LLM-based approach because they are commonly complemented with other tools or tester experience. For generating assertions (Fig.~\ref{fig:approaches_to_generate_oracles}~\textcircled{A}), the LLM receives the program/test information and generates an oracle in the form of an assertion, post-condition, or invariant. This assertion is assembled with the test prefix to obtain a full test case. Next, the test case can be improved by incorporating the libraries/packages and solving minor syntax issues, e.g., adding parentheses when it is not closed. Finally, the test case is compiled and executed to detect issues in the assertion. If the test has some compilation/execution issues, the LLM could be prompted again to refine the test case based on the error messages.

On the other hand, the LLM-based approaches aimed at deriving and generating metamorphic relationships follow the process of Fig.~\ref{fig:approaches_to_generate_oracles}~\textcircled{B}. First, the LLM only receives the program specification and derives several metamorphic relationships. If they are not good enough, the tester again prompts the LLM to provide feedback to enhance them. Once the metamorphic relationships are derived, the LLM generates executable code considering both the program under test and some examples.

The third group of articles related to oracles (Table~\ref{tab:types_oracles}) is about reflections. Molina et al. ~\cite{Molina2024} survey 37 articles from 2020 to 2024 on LLMs and oracles. Note that our study is broader, and we classify some of these 37 articles in other categories. For example, we classify as \textit{``Unit Test Generation''} (Section~\ref{sec:llmusage:unittestgen}) those articles about test completion that are primarily focused on generating the test structure rather than addressing the assertions. In the survey of Molina et al.~\cite{Molina2024}, they analyze both the challenges and the future direction. Among them, one challenge is the capability of the LLM-based approach to detect bugs. Zhang et al.~\cite{ZhangTSE2025} conduct an extensive study concluding that LLM-based approaches outperform the other state-of-the-art techniques to generate oracles. They also analyze the potential of using fine-tuning. In this direction, Yibo et al.~\cite{Yibo2024} study the focal methods and propose a dataset that can be used to fine-tune models. Evaluating LLM-based approaches is challenging, and Jiho et al.~\cite{Jiho2024} compare the common metrics used. They found a weak relationship between the textual metrics (e.g., BLEU) and the test adequacy metrics (e.g., mutation score), so the authors do not recommend using textual metrics to evaluate the effectiveness of the oracles. Another issue with the evaluation is the use of inappropriate settings. Soneya et al.~\cite{Soneya2023} discover that some faults are triggered just by the exceptions of test prefix code, regardless of the oracle generated. If this is not controlled in the experiments, it can overestimate the fault capability. To this end, Zhongxin et al.~\cite{Zhongxin2023} propose to use a baseline called NoException to check that the test prefix does not raise an exception.

Several articles consider TOGA~\cite{Elizabeth2022} and ATLAS~\cite{Watson2020} as the state-of-the-art and use them as a baseline. On the one hand, TOGA outperforms AthenaTest~\cite{Tufano2021} and other pre-trained BART~\cite{Tufano2022} in terms of bugs detected in Defects4J. At the same time, TOGA is outperformed by several other approaches:
\begin{itemize}
    \item Approaches that caught more bugs: n2postcond ~\cite{Madeline2024} and TOGLL~\cite{Hossain2025}.
    \item Approaches that improve the success rate: AugmenTest~\cite{Khandaker2025}.
    \item Approaches that have more accuracy and BLEU similarities: AssertT5~\cite{Primbs2025}, TECO~\cite{Nie2023} and ChatAssert~\cite{Hayet2025}.
\end{itemize}

Despite outperforming TOGA, some of them are complementary because they detect different faults~\cite{Madeline2024}. On the other hand, ATLAS~\cite{Watson2020} is another LLM-based approach that is commonly used as a baseline. The approaches CEDAR~\cite{Noor2023} and ~\cite{ZhangTSE2025} improve the accuracy. The AssertT5~\cite{Primbs2025}, TECO~\cite{Nie2023}, and ChatAssert~\cite{Hayet2025} not only improve the accuracy but also the BLEU. Finally, ~\cite{Hailong2024} improves both accuracy and bug detection.

Other articles do not use TOGA~\cite{Elizabeth2022} or ATLAS~\cite{Watson2020} as a baseline. A3Test~\cite{Alagarsamy2024} generates better asserts than AthenaTest~\cite{Tufano2021} and ChatUnitTest~\cite{ChenFSE2024}. At the same time, A3Test~\cite{Alagarsamy2024} is also outperformed in terms of accuracy by~\cite{Ni2024}, which also outperforms CAT-LM~\cite{Rao2024}.

The common metrics used to evaluate the assertions can be divided into those related to text similarity or test adequacy. These metrics are commonly compared with a ground truth oracle created by a human. The exact match accuracy compares if both are equal, but ROUGE, METEOR, and BLEU metrics compute how similar/far apart they are. Another metric is EditSim, which measures the number of changes required in the oracle to obtain the ground truth. On the other hand, the adequacy metrics most used are the number/percentage of bugs found and the mutation score. Some articles also measure the execution time, the coverage, and the success rate of compilation/execution.

The benchmarks commonly used in the evaluation are Defects4J, Methods2Test, ATLAS, and CodeSearchNet. However, it is also common to evaluate the assertions using custom benchmarks from GitHub or other repositories.

\subsection{Test Augmentation or Improvement}
\label{sec:llmusage:augmentation}

The common goal of the works in this category is to leverage LLMs to enhance existing test suites for benchmarks or industrial applications. We consider that such enhancements are achieved by following several strategies: augmenting the existing test suites, or leading to some improvements on the overall testing processes.

\begin{figure}[htb]
 \centering
 \includegraphics[width=0.9\textwidth,trim=60 160 60 160, clip]{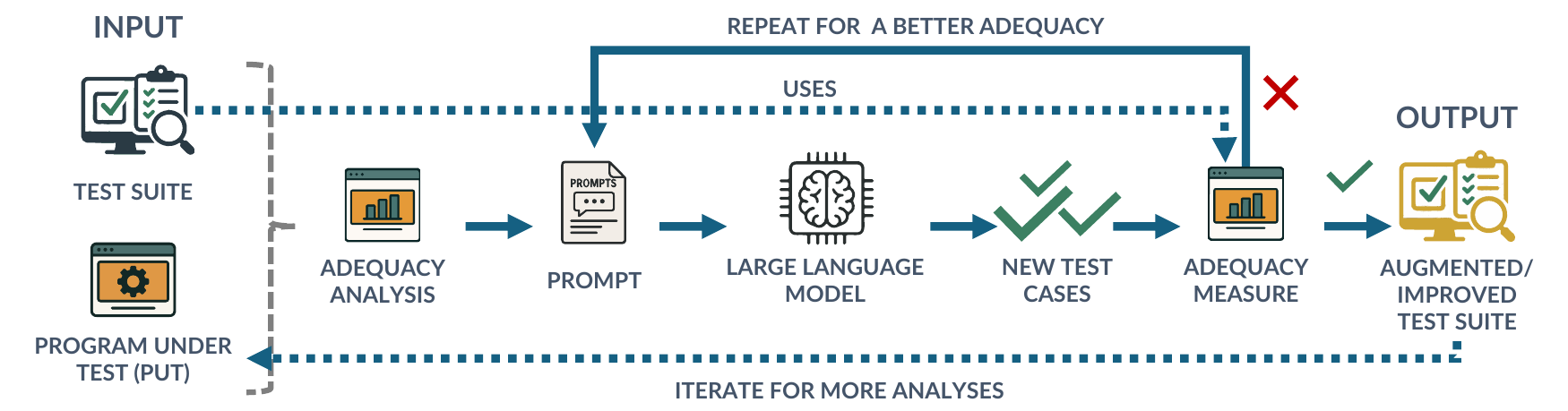}
 \caption{Test Augmentation or Improvement Process}
 \label{fig:augmentation}
\Description[Test Augmentation or Improvement Process]{Test Augmentation or Improvement Process}
\end{figure}

Fig.~\ref{fig:augmentation} depicts a high-level interaction schema for test augmentation approaches based on LLMs. The schema is iterative, and is guided by an adequacy analysis based on some quality attributes.
Overall, the process can be summarized as follows: first, the quality of the current test suite is evaluated against the considered PUT; the adequacy analysis on the observed results guides the engineering of the prompts that will be submitted to the LLM. The expected outcome of the interaction with the LLM is a collection of new test cases; thus, their quality is evaluated (possibly also taking into account the original test suite). These few steps are repeated iteratively until the expected adequacy criteria are met. The resulting augmented test suite is then used to verify the PUT, and possibly it is further improved by leveraging additional iterations.

Under this interaction schema, the adequacy analysis could refer to several (alternative) quality criteria. For example,~\cite{Alshahwan2024} uses coverage information; while others, like~\cite{Dakhel2024}, refer to fault detection capabilities based on mutants. In both cases, adequacy analysis results can be used in order to prompt LLM for the generation of test cases that cover a specific and not yet considered function or portion of the PUT, as in~\cite{Altmayer2025}. 

In other works, LLM is used as a complement of existing automatic test case generation approaches; for example, in order to overcome their known issues when a local optimum for the considered adequacy metric is reached. Specifically,~\cite{Lemieux2024} leverages a traditional SBST approach until its coverage improvements stall, then the approach invokes an LLM to provide test cases addressing under-considered functions. These new test cases are used as seeds for the subsequent iterations on SBST. Similarly,~\cite{Meng2024} generates test inputs by means of a grey-box fuzzing approach, and when it does not succeed in increasing coverage anymore, LLM is prompted to generate a new input that could overcome any coverage plateau.

Some approaches in test augmentation operate autonomously or with limited human interactions; others instead explicitly foresee humans-in-the-loop. For example, in~\cite{Paduraru2024}, humans collect/define sets of test cases for games built on specific game engines. These tests are also used to fine-tune a code-oriented LLM. The resulting model is then used to generate additional unit test cases. In this specific work, the generation process is mainly guided by coverage improvements. Still, it also makes use of RAGs in order to keep the synthesized unit tests aligned with specific development practices.

Works in test improvement investigates how the adoption of LLMs can impact established software testing practices. For example,~\cite{WangArxiv2024} studies how to improve mutation testing by means of mutants that are closer to real bugs by developers. Similarly, some works support the generation of mutants for domain-specific languages (e.g., Simulink models in~\cite{ZhangAST2025}) that, differently from canonical mutants based on mutation patterns, are easier to trace against requirement violations. 

LLMs can also contribute to the improvement of the test process. For example, they have been used in order to support the evolution of the artifacts (e.g., specifications, test code) related to a given PUT. In this direction,~\cite{Myeongsoo2024} proposes to enhance the OpenAPI specifications of REST services with information (e.g., data constraints, examples, and rules) returned by interactions with an LLM. There, the prompting is guided by data explicitly available in both machine-readable and human-readable sections of the original OpenAPI specification. Differently, there are also works that deal with the co-evolution of test cases and the target PUT: for example,~\cite{LiuISSRE2024} LLM exploits a given contextual information from a PUT or its usage in order to update potentially obsolete test cases automatically.

Flakiness is commonly referred to as the intermittent pass and failure outcomes exhibited by the same set of test cases despite exercising the unchanged code~\cite{bbd:2021}. Understanding the root cause of flakiness is a major problem that has been deeply investigated during the last few years. LLMs have also been used to parse a test case implementation to predict whether it is flaky or not. As reported in~\cite{Rahman2024}, the foreseen advantage is that (fine-tuned) LLMs can achieve good classification accuracy while avoiding the need to re-run the tests several times. Notably, these works concern test improvements as their main focus is on the test itself and not in possible operating conditions that could eventually be experienced in the field~\cite{bbd:2024}. Once a specific flaky test has been disclosed, LLMs can also be used in order to guide its evolution based on the identified category of flakiness (e.g., like in~\cite{Fatima2024}).

\subsection{Test Agents}
\label{sec:llmusage:testagents}

\begin{figure}[htb]
\centering
\includegraphics[width=0.9\textwidth,trim=100 120 100 120, clip]{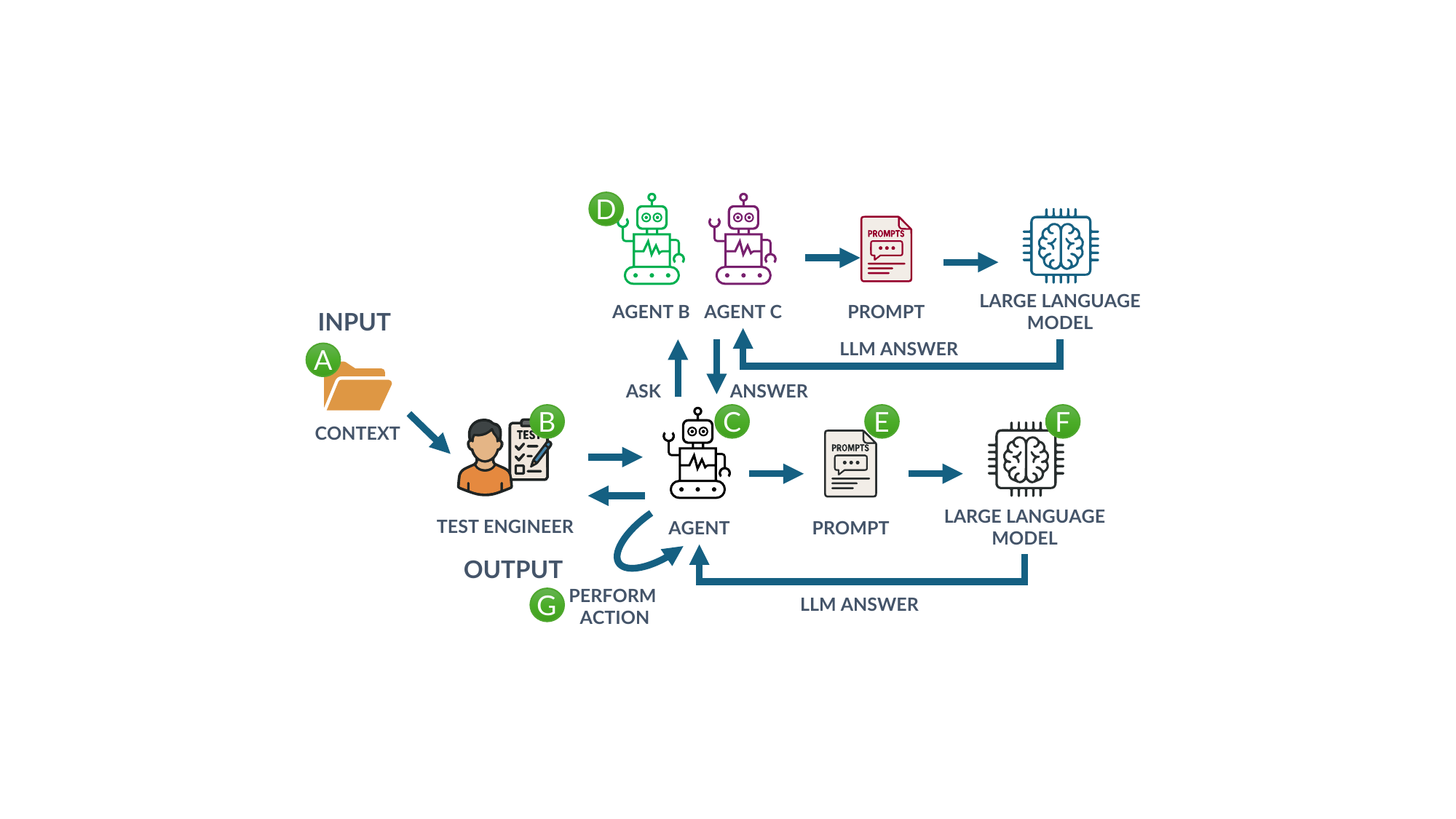}
\caption{LLM based Testing Test Agents Process} \label{fig:agentsprocess}
\Description[LLM based Testing Test Agents Process]{LLM based Testing Test Agents Process}
\end{figure}

Test agents are lightweight clients that interact with the user and automate testing tasks, like the test suite configuration and execution~\cite{Bouzenia2024}, the test generation~\cite{Garlapati2024, Yoon2024, Jorgensen2024, Almutawa2024}, security testing~\cite{Bianou2024, Wu2024}, or serve as an interface between humans and a broad range of testing tasks~\cite{Feldt2023}.
Fig.~\ref{fig:agentsprocess} depicts the process of LLM-based testing agents. Test agents are characterized by having the \textit{``tester in the loop''}, which is the individual who initiates the first interaction or performs tasks with the agent.  
The test engineer gets the context to the agent (Fig.~\ref{fig:agentsprocess}~\textcircled{A}-\textcircled{B}), asking the agent to perform some task, for example, with a reference to a project or its URI~\cite{Bouzenia2024, Wu2024}, or initiate some process~\cite{Bianou2024}. 
In the reviewed articles, the test agent (Fig.~\ref{fig:agentsprocess}~\textcircled{C}) can prompt the LLM~\cite{Feldt2023, Bouzenia2024, Yoon2024, Jorgensen2024, Wu2024, Almutawa2024}, using custom prompts and retrieve the answers. 
Another work~\cite{Garlapati2024, Bianou2024} proposes to employ the most suitable agent, redirecting the request (Fig.~\ref{fig:agentsprocess}~\textcircled{D}) and waiting for its response.

The answer of the LLM or agent can be redirected to the user~\cite{Feldt2023, Wu2024}, or derivative into a specific actions (Fig.~\ref{fig:agentsprocess}~\textcircled{G})  like configure and execute the test suite or other systems~\cite{Feldt2023, Jorgensen2024}, generate test cases~\cite{Garlapati2024, Yoon2024, Almutawa2024}, check the non functional quality of a system~\cite{Bianou2024}.

The evaluation of those agents uses context-dependent benchmarks (e.g., mobile test cases are performed against the Themis app benchmark) and metrics (e.g., when generating test cases, employ the percentage of test cases that pass or build, whereas in security tests, employ the successful exploits). 
Most works employ GPT family models, open-source models like Llama, and code-purpose models like CodeLlama.

\subsection{Non-Functional Testing}
\label{sec:llmusage:nofunctest}
\begin{figure}[htb]
\centering
\includegraphics[width=0.9\textwidth,trim=60 180 60 180, clip]{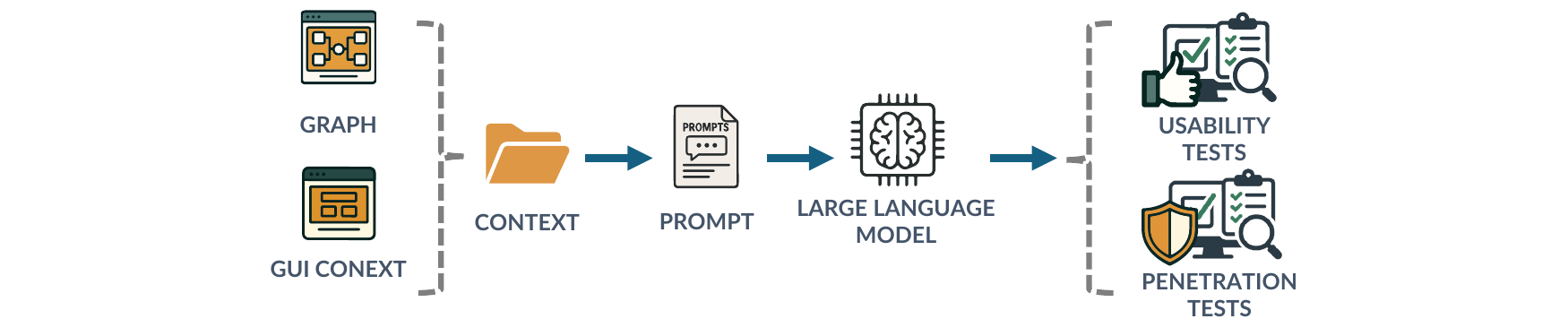}
\caption{LLM-based Non-functional testing process} \label{fig:nonfuncprocess}
\Description[LLM-based Non-functional testing Process]{LLM-based Non-functional testing process}
\end{figure}

The Non-functional LLM-based testing category gets \numnonfuct articles, mainly focused on security and usability testing. Some of these works follow the test agent process  (See Fig.~\ref{fig:agentsprocess})~\cite{Wu2024, Bianou2024, Happe2023}, the high-level test generation with the LLM in the loop  (See Fig.~\ref{fig:highendprocess})~\cite{Chen2024}, whereas other follow a simple process depicted in the Fig.~\ref{fig:nonfuncprocess}.

The security articles use as input the system specifications, like OpenAPI~\cite{Pasca2025}, that are augmented with API fragments and historical test data, state graph diagrams~\cite{SuQingram2024}, scrape the context and UI of applications~\cite{Chen2024}, or context provided by the tester. These works use a general-purpose LLM~\cite{Wu2024,Deng2024, Bianou2024, Pasca2025, Calvano2025, Happe2023} but also approaches fine-tune the model with requirements and examples~\cite{SuQingram2024,Chen2024}.
The usability article retrieved employs the UI context~\cite{Calvano2025} as input to prompt the LLm and retrieve as output different tasks to perform in usability testing.

In both security and usability, the models are prompted with the necessary context, following the strategy depicted in the different processes, and providing as output the different steps to perform an action (e.g., an attack)~\cite{Happe2023}, several test cases~\cite{Pasca2025, SuQingram2024}, and executing the action~\cite{Deng2024, Wu2024}.

The approaches are evaluated using metrics and benchmarks specific to their field and the desired outcomes. Most utilize LLMs from the GPT family, as well as those from the Llama family, Mistral, Gemini, or Qwen 2.

%% file: sheet3_1.tex
\begin{table}[hbt]
  \caption{Types of LLM-based testing articles about oracles}
  \label{tab:types_oracles}
  \centering
  \begin{adjustbox}{max width=\linewidth}
  \begin{tabular}{m{17.355em}m{20.355em}>{\centering\arraybackslash}m{5.355em}}
    \toprule
   \textbf{ Types of articles related to oracles} & \textbf{Articles} & \textbf{TOTAL}\\
    \midrule
    Generate assert & \textbf{P35}~\cite{Khandaker2025}, \textbf{P49}~\cite{Tufano2022}, \textbf{P50}~\cite{Watson2020}, \textbf{P51}~\cite{Primbs2025}, \textbf{P62}~\cite{Konstantinou2024}, \textbf{P63}~\cite{Hayet2025}, \textbf{P64}~\cite{Hossain2025}, \textbf{P67}~\cite{Noor2023}, \textbf{P75}~\cite{Elizabeth2022}, \textbf{P76}~\cite{Nie2023}, \textbf{P81}~\cite{Alagarsamy2024}, \textbf{P85}~\cite{Hailong2024}, \textbf{P87}~\cite{Ni2024}, \textbf{P93}~\cite{Madeline2024}, \textbf{P144}~\cite{Huynh2024}  & 15/26 (58\%) \\
    \midrule
    Generate metamorphic relationships & \textbf{P89}~\cite{Hung2023}, \textbf{P90}~\cite{Yifan2023}, \textbf{P91}~\cite{Seung2024}, \textbf{P92}~\cite{Congying2024}, \textbf{P145}~\cite{TsigkanosICCS2024} & 5/26 (19\%) \\
    \midrule
    Reflections & \textbf{P23}~\cite{Molina2024}, \textbf{P77}~\cite{Soneya2023}, \textbf{P79}~\cite{Jiho2024}, \textbf{P80}~\cite{Zhongxin2023}, \textbf{P82}~\cite{Yibo2024}, \textbf{P84}~\cite{ZhangTSE2025} & 6/26 (23\%) \\
\bottomrule
  \end{tabular}
  \end{adjustbox}
\end{table}

%% file: sheet3_2.tex
\begin{table}[htb]
    \centering
    \caption{Inputs used by LLM-based approaches to generate oracles}
    \label{tab:types_of_prompt_inputs_for_oracles}
    \adjustbox{max width=\textwidth}{
        \begin{tabular}{m{2em}r m{18em} m{4.5em}c }
            \toprule
            \multicolumn{2}{p{11em}}{\textbf{LLM input}} & \textbf{Articles} & \multicolumn{2}{p{8.785em}}{\textbf{TOTAL}} \\
            \midrule
            \multicolumn{1}{r}{\multirow{3}[6]{*}{Program Under Test(PUT)}} 
                & \multicolumn{1}{p{9em}}{Focal method} 
                & \textbf{P49}~\cite{Tufano2022}, \textbf{P50}~\cite{Watson2020}, \textbf{P51}~\cite{Primbs2025}, \textbf{P62}~\cite{Konstantinou2024}, \textbf{P63}~\cite{Hayet2025}, \textbf{P64}~\cite{Hossain2025}, \textbf{P67}~\cite{Noor2023}, \textbf{P76}~\cite{Nie2023}, \textbf{P81}~\cite{Alagarsamy2024}, \textbf{P85}~\cite{Hailong2024}, \textbf{P92}~\cite{Congying2024}, \textbf{P93}~\cite{Madeline2024} 
                & \multicolumn{1}{p{4.5em}}{12/20 (60\%)} 
                & \multicolumn{1}{c}{\multirow{3}[6]{*}{15/20 (75\%)}} \\
            \cmidrule{2-4} 
            \multicolumn{1}{r}{} 
                & \multicolumn{1}{p{9em}}{Focal method signature} 
                & \textbf{P35}~\cite{Khandaker2025}, \textbf{P75}~\cite{Elizabeth2022},           
                \textbf{P76}~\cite{Nie2023}, \textbf{P87}~\cite{Ni2024} 
                & \multicolumn{1}{p{4.5em}}{4/20 (20\%)} 
                &  \\
            \cmidrule{2-4} 
            \multicolumn{1}{r}{} 
                & \multicolumn{1}{p{9em}}{Other methods or class} 
                & \textbf{P35}~\cite{Khandaker2025}, \textbf{P62}~\cite{Konstantinou2024}, \textbf{P63}~\cite{Hayet2025}, \textbf{P76}~\cite{Nie2023}, \textbf{P94}~\cite{Kexin2023} 
                & \multicolumn{1}{p{4.5em}}{5/20 (25\%)} 
                &  \\
            \midrule
            \multicolumn{2}{p{11em}}{Test prefix} 
                & \textbf{P35}~\cite{Khandaker2025}, \textbf{P49}~\cite{Tufano2022}, \textbf{P50}~\cite{Watson2020}, \textbf{P51}~\cite{Primbs2025}, \textbf{P62}~\cite{Konstantinou2024}, \textbf{P64}~\cite{Hossain2025}, \textbf{P67}~\cite{Noor2023}, \textbf{P75}~\cite{Elizabeth2022}, \textbf{P76}~\cite{Nie2023}, \textbf{P85}~\cite{Hailong2024}, \textbf{P87}~\cite{Ni2024}, \textbf{P92}~\cite{Congying2024} 
                & \multicolumn{2}{p{8.785em}}{12/20 (60\%)} \\
            \midrule
            \multicolumn{1}{l}{\multirow{3}[6]{*}{Context}} 
                & \multicolumn{1}{p{9em}}{Documentation comments (e.g. docstring)} 
                & \textbf{P35}~\cite{Khandaker2025}, \textbf{P62}~\cite{Konstantinou2024}, \textbf{P64}~\cite{Hossain2025}, \textbf{P75}~\cite{Elizabeth2022} 
                & \multicolumn{1}{p{4.5em}}{4/20 (20\%)} 
                & \multicolumn{1}{c}{\multirow{3}[6]{*}{11/20 (55\%)}} \\
            \cmidrule{2-4} 
            \multicolumn{1}{r}{} 
                & \multicolumn{1}{p{9em}}{Program specification} 
                & \textbf{P89}~\cite{Hung2023}, \textbf{P90}~\cite{Yifan2023}, \textbf{P91}~\cite{Seung2024}, \textbf{P93}~\cite{Madeline2024}, \textbf{P144}~\cite{Huynh2024}, \textbf{P145}~\cite{TsigkanosICCS2024} 
                & \multicolumn{1}{p{4.5em}}{6/20 (30\%)} 
                &  \\
            \cmidrule{2-4} 
            \multicolumn{1}{r}{} 
                & \multicolumn{1}{p{9em}}{Dependencies/types} 
                & \textbf{P35}~\cite{Khandaker2025}, \textbf{P76}~\cite{Nie2023} 
                & \multicolumn{1}{p{4.5em}}{2/20 (10\%)} 
                &  \\
            \midrule
            \multicolumn{2}{p{11em}}{Examples} 
                & \textbf{P63}~\cite{Hayet2025}, \textbf{P67}~\cite{Noor2023}, \textbf{P145}~\cite{TsigkanosICCS2024} 
                & \multicolumn{2}{p{8.785em}}{3/20 (15\%)} \\
            \midrule
            \multicolumn{2}{p{11em}}{Other} 
                & \textbf{P92}~\cite{Congying2024} 
                & \multicolumn{2}{p{8.785em}}{1/20 (5\%)} \\
            \bottomrule
        \end{tabular}%
    }
    \label{tab:addlabel}%
\end{table}%

%% file: 06_reflection.tex
\section{LLM-based testing research challenges and foreseen impact}
\label{sec:reflections}

In the following, we analyze the challenges and opportunities that LLMs are leading to software testing research. Specifically, our analysis is organized in both Section~\ref{sec:reflections:roadmap} and Section~\ref{sec:reflections:dreams}, which respectively address \textbf{RQ3} and \textbf{RQ4}. 

\subsection{Roadmap of LLM-based testing research}
\label{sec:reflections:roadmap}

\begin{figure}[htb]
\centering
\includegraphics[width=0.9\textwidth,trim=0 0 0 0, clip]{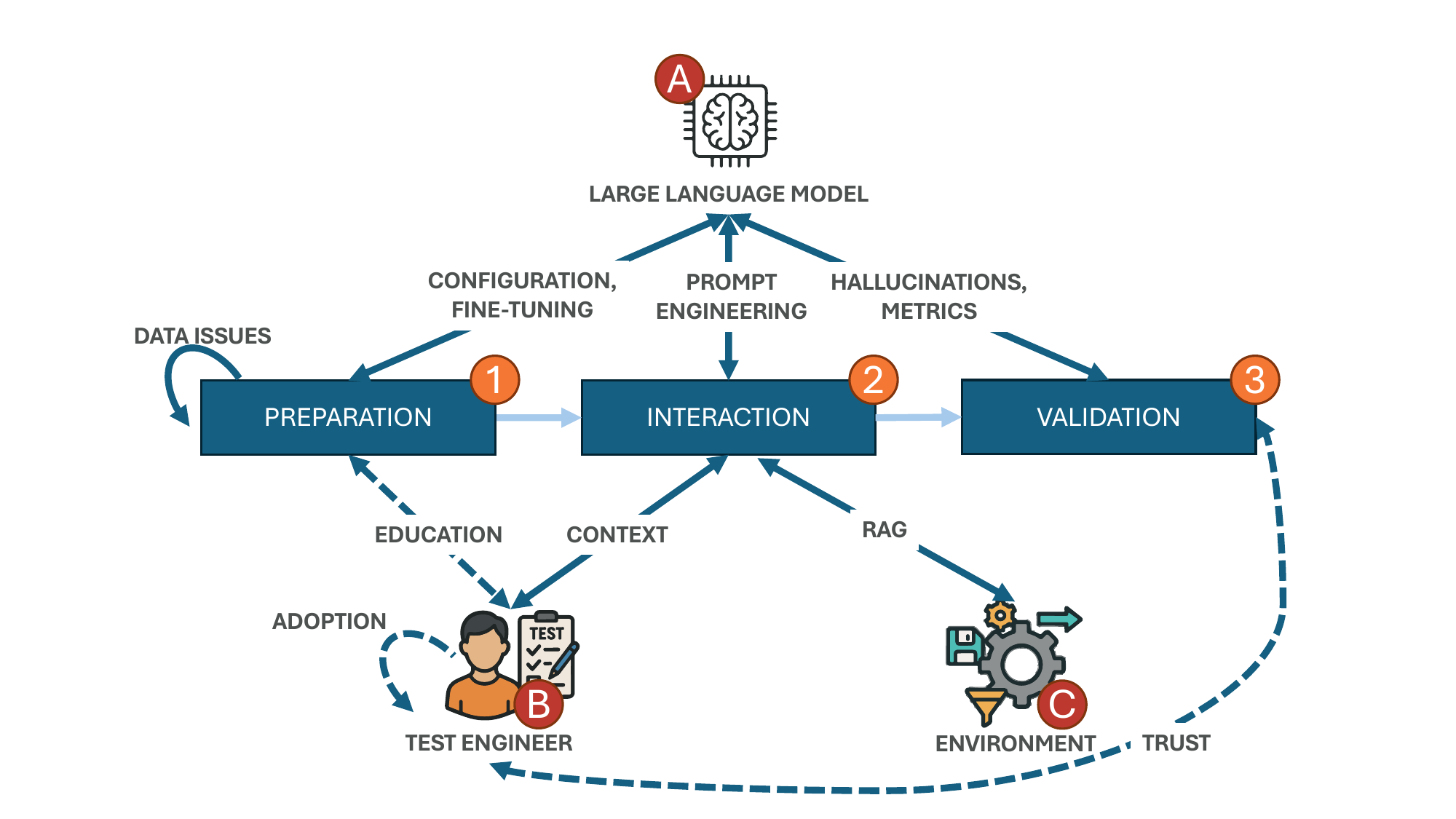}
\caption{LLM-based software testing conceptual framework} \label{fig:reflconcep}
\Description[LLM-based software testing conceptual framework]{LLM-based software testing conceptual framework}
\end{figure}

Based on our overview of articles on LLM-based software testing, we propose a research roadmap in this section. 
Fig.~\ref{fig:reflconcep} depicts a conceptual framework that models the main actors covered by our roadmap and the stages where they can be involved.

Precisely, the actors represent the main entities that cooperate during the LLM-based testing activity. Namely they are: \textcircled{A} the LLM, \textcircled{B} the test engineer (or simply tester), and \textcircled{C} the environment referred during the interactions. 
We also consider three main stages, represented in Fig.~\ref{fig:reflconcep} as blue boxes: 
Preparation, Interaction, and Validation.
The \textit{Preparation} stage (Fig.~\ref{fig:reflconcep}~\textcircled{1}) refers to the setup activities that are performed before any cooperation among actors start.
Next, the \textit{Interaction} stage  (Fig.~\ref{fig:reflconcep}~\textcircled{2}) concerns activities where different actors are requested to cooperate by affecting each other.
Finally, the \textit{Validation} stage (Fig.~\ref{fig:reflconcep}~\textcircled{3}) checks that both the achieved outcomes, and the exposed behavior conforms with what expected.
Each arrow in Fig.~\ref{fig:reflconcep} models a specific research challenge we identified, and that affects a given actor at a given stage. Continuous arrows denote challenges related to technical aspects, while dashed ones represent social challenges related to human factors.

In the following, we discuss all these challenges in detail: at first, by reporting technical challenges per stage (i.e., Section~\ref{sec:preparation}, Section~\ref{sec:interaction}, and Section~\ref{sec:validation}); then we outline the social challenges we encountered (i.e., Section~\ref{sec:social}).

\subsubsection{Preparation}
\label{sec:preparation}
The preparation stage involves configuring the models, such as setting their parameters appropriately, fine-tuning them for improved performance in the specific testing context, and handling the data used for training and testing. Although general-purpose LLMs with standard configurations demonstrate promising capabilities in LLM-based testing, three key challenges arise to improve their performance and results: (1) selecting the optimal configuration settings, (2) how to effectively fine-tune the models, and (3) how to manage the data issues to ensure a fair and reliable evaluation. 
In detail:

\begin{itemize}
    \item \textbf{Configuration:}  although promising results have been obtained with specific configurations, several authors have explicitly identified exploring more configurations as an open challenge.
    Specifically, some researchers have pointed out the need for more studies across a broad range of models~\cite{Jiang2024, Augusto2025, Gao2025, Meng2024, Nie2023, Happe2023, SantosSBQS2024, TsigkanosICCS2024, WangICSME2024, XiaoInt2024}. In contrast, others have focused on the parametrization of their LLM model~\cite{Pasca2025, Taherkhani2024}. For instance, the influence of settings such as temperature values and window sizes has been noted as an area that requires further research~\cite{Pasca2025}. Similarly, the impact of hyperparameter tuning on model performance represents another direction that requires investigation~\cite{Taherkhani2024}.
    A related research challenge is that of evaluating the use of LLMs within more programming languages or testing frameworks~\cite{Gao2025, ChenArxiv2024, Meng2024, Liu2023, Tufano2021, Elizabeth2022, Nie2023, Hailong2024}. Some studies as in~\cite{Bouzenia2024} report that LLMs perform well when they are used to test software in some languages but not well in others. 
    Other studies~\cite{Taherkhani2024} hint that language-related performance degradation could be mitigated by means of minimal fine-tuning.
    
    \begin{acmmdframed}
        \textit{Understanding the interactions between different configurations and programming languages, including parameter tuning and the selection of different models, and their impact on outcomes, remains an open challenge in LLM-based testing.}
    \end{acmmdframed}

    \item \textbf{Fine-Tuning:} research about fine-tuning in LLM-based testing is still ongoing. While fine-tuned models have been explored in all categories with promising results (Fig.~\ref{fig:yearsandcat}-B, red and pink-red colors), several challenges remain unaddressed.
    Several articles have identified fine-tuning as a key research frontier for LLM-based testing~\cite{Plein2024, Yu2023, ChenArxiv2024, LiuGUI2024, Pan2025, Kang2024, TsigkanosICCS2024, Yin2024, PaduraruJournal2024, Petrovic2024, Munley2024, Bianou2024}. On one hand, several authors anticipate future improvements from refining and fine-tuning the model in their LLM-based approaches~\cite{Yin2024, Kang2024, Schwachhofer2024, Lemieux2024, Happe2023, Munley2024}. Still, on the other hand, it is commonly acknowledged~\cite{Luo2025} that incorrect fine-tuning can harm model performance and cause catastrophic forgetting, i.e.,  the model loses previously acquired knowledge.
    The most decisive factor is often the data used during fine-tuning: high-quality and relevant input data remains an acknowledged challenge by several authors~\cite{Hoffman2024, Yu2023, Zheng2025, Petrovic2024}.
    Those works also acknowledge that if the data collection is curated manually (e.g., for its annotation) or requires domain-specific knowledge~\cite{XueISSTA2024}, the cost of developing the model increases dramatically.
    On the performance side, some findings suggest that smaller fine-tuned models can outperform larger general-purpose LLMs~\cite{PaduraruJournal2024, Pan2025}.
    This competitive advantage has drawbacks: fine-tuning typically requires access to specialized hardware or cloud infrastructure~\cite{Shang2024, Rao2024}, and specific knowledge is not always present in testing teams.
    One research direction to reduce the resources used in LLMs is to use the computation employed to gain knowledge for solving one task (e.g., one programming language) to solve a different but related task~\cite{Liu2023, Elizabeth2022, Nie2023, Hoffman2024, ZhangICSE2025, ChenArxiv2024}. This means exploring the applicability of fine-tuned models in one context to another.

    \begin{acmmdframed}
        \textit{The trade-off in fine-tuning involves balancing precision, cost, and flexibility, leading to future research opportunities. The research in this challenge could investigate whether to develop specialized tools for specific tasks or general-purpose models that adapt across various contexts through the use of prompt engineering.}
    \end{acmmdframed}

\end{itemize}

\begin{itemize}
    \item \textbf{Data Issues:}
    a key challenge in LLM-based testing lies in the data leakage problem, widely acknowledged in literature~\cite{Alshahwan2024, OuedraogoArxiv2024, Molina2024, Jiang2024, Abdullin2025, Meng2024, Hayet2025, Rao2024}.
    Traditionally, the common evaluation approaches rely on splitting the data between training and validation to ensure a fair evaluation. 
    Where training data is available or reproducible, one growing trend is replicating original train-validation splits during testing~\cite{Nie2023}.
    However, with most commercial LLMs, this is impossible because they do not disclose the datasets used during training, hindering the evaluation with truly \textit{``unseen''} data~\cite{WangTSE2024}. Validating the model with the same data employed in development can cause a model that excels in development, but it underperforms in production when faced with unseen data.
    \textit{ This opacity undermines the reliability of LLMs and raises serious concerns about reproducibility and trust.}
     Still, due to opacity,  a tester asking LLM to generate a test suite could unknowingly undergo the risk of violating intellectual property~\cite{Samuelson2023}, if the LLM returns code protected by proprietary rights. 
    In response to the above data issues, several researchers have acknowledged the need for new, diverse, and standardized benchmarks~\cite{Savage2024, OuedraogoArxiv2024, Molina2024, Abdullin2025, Alian2025, Wang2024}, leading to the release of new datasets and evaluation benchmarks~\cite{WangTESTEVAL2024, Lops2024, He2024}.
    Some authors propose evaluating the approaches using custom or private datasets to minimize data leakage risks~\cite{Abdullin2025}. However, in doing this, we should be aware of potential security threats, as it is well known that LLMs could release sensitive or private data~\cite{yan2025protecting}. In contrast, others suggest evaluating LLM-based testing approaches using older models, assuming a new benchmark or dataset was not used during their training.
    This creates a treadmill effect: as models evolve, so must the benchmarks, leading to an endless race that exhausts academia and industry.

    \begin{acmmdframed}
    \textit{As research on LLM-based testing matures, new aspects are becoming central to the reliability and reproducibility of its results.
    Among the others, the literature fosters investigation about managing data leakages, dataset privacy, intellectual property rights, and fairness of the computational process.}
    \end{acmmdframed}

    \end{itemize}

\subsubsection{Interaction}
\label{sec:interaction}

Let us now consider a model instance that has already been configured and that is ready to be used. We identified several challenges related to how test engineers guide interactions with the LLM. Specifically, the literature highlighted the strong influence between the way a request is phrased and the outcome an LLM produces. Similarly, it appears quite central the way test engineers express additional contextual information in order to effectively achieve some goal. Finally, it is also important to consider how to retrieve external knowledge and how to dynamically support the LLM in order to integrate it. The detailed presentation of these challenges is reported below:

\begin{itemize}
    \item \textbf{Prompt Engineering:} refers to the process of guiding the generative AI systems to produce the desired outputs\footnote{https://aws.amazon.com/what-is/prompt-engineering}. A growing body of research, both general~\cite{Zamfirescu2023} and testing-specific~\cite{Schafer2024, Lemieux2024, WangTESTEVAL2024, Jiang2024, OuedraogoASE2024, Lops2024, Fatima2024, Lin2024, LiuGUI2024, Ni2024, Happe2023, Wang2024, Jiri2024, XiaoInt2024, TsigkanosICCS2024, ChenArxiv2024}, has highlighted the important role of prompt crafting in the LLMs' performance.
    \textit{The challenge lies in systematically designing prompts that guide the model toward useful, valid, and in-context outputs.}
    To address this challenge, several promising directions have emerged. One line of future work proposes to explore multi-prompt strategies, aiming to improve the model performance by diversifying the prompt input~\cite{Feldt2023}. Others have proposed employing \textit{in-context learning}, enhancing prompt accuracy by including representative examples directly within the input~\cite{Fatima2024}. Others have proposed to employ \textit{in-memory programming}, where the previous LLM output is included in the next interaction~\cite{Happe2023}. Another direction is the so-called \textit{self-refinement}, where the LLM is prompted to enhance the relevance and the clarity of the prompt~\cite{Wang2024}. \textit{These strategies reflect a growing interest in becoming the prompt design into a non-static, iterative, and adaptive process}. Some authors have shown that with well-crafted prompts, general models can reach comparable performance~\cite{Shang2024, Munley2024, Liu2022} to fine-tuned models. These challenges emphasize the need to introduce \textit{Promptware}~\cite{Chen2025} testing approaches to address prompt design and validation, aiming for more reliable, maintainable, and testable LLM-based testing methods.
    \begin{acmmdframed}
    \textit{
    Prompt engineering and prompt quality attracted a growing interest in the software testing community working with LLMs. Despite this, there is still room for innovative methods to evaluate, validate, and improve prompt quality in this domain.}
      
    \end{acmmdframed}
\end{itemize}
\begin{itemize}
    \item \textbf{Context:} refers to the information provided to the model to perform a specific task. Several works have emphasized the potential of enriching it to improve the performance~\cite{ChenFSE2024, Huang2024, Yu2023, Noor2023, Tufano2021, WangICSME2024, Yin2024}.
    \textit{The challenge is how to provide a broader base of relevant testing knowledge to enhance the LLM performance.}
    However, some authors also highlight that more context does not always guarantee better performance~\cite{PaduraruJournal2024}. Increasing the context is sometimes counterproductive, reducing the model's accuracy~\cite{PaduraruJournal2024}. Some authors have also acknowledged the challenge of limited context due to input constraints~\cite{Deng2024}, or to automate the data selection and extraction process~\cite{Lops2024, Primbs2025}. 
    \textit{The question of how much context is enough—and which context is effective—remains largely open.}. Another common target for many authors is the context quality. It is achieved by improving the input~\cite{Chetan2024, SantosSBQS2024}, integrating domain knowledge~\cite{XueISSTA2024}, or enhancing how the information is provided~\cite{WangTESTEVAL2024, Yin2025, ChenArxiv2024, Noor2023, Deng2024}.

    Another author's target for future work is not the quantity of context, but its quality, focusing on improving the input~\cite{Chetan2024, SantosSBQS2024} by integrating domain knowledge~\cite{XueISSTA2024}, or improving how the information is provided~\cite{WangTESTEVAL2024, Yin2025, ChenArxiv2024, Noor2023, Deng2024}. \textit{These works recognize that the efforts should be put not only in the amount of context but also in its quality and the alignment with the test objectives.}
    Instead of assuming high-quality inputs, other works propose improving the model's performance even in minimal, irrelevant,  or noisy context scenarios~\cite{Yang2025}. 
    \textit{This approach underscores the need for LLMs to be robust under real-world constraints, where ideal inputs cannot always be guaranteed.}

    \begin{acmmdframed}
        \textit{The relationships between context size, quality, and model robustness outline a rich and broad research landscape where trade-offs must be carefully negotiated.}
    \end{acmmdframed}

\end{itemize}
\begin{itemize}
    \item \textbf{Retrieval Augmented Generation (RAG):}  has been explored in several LLM-based software testing studies to enrich the input context provided to the model~\cite{Munley2024, Alagarsamy2024, Chetan2024, Wanigasekara2024, Khandaker2025}.
    Some works have highlighted that despite expanding the model's input window, one of the challenges of RAG is addressing those cases where the retrieval context exceeds the model's processing capacity~\cite{Chetan2024, Wanigasekara2024}. On the other hand,  RAG is also used to address those long contexts within the prompt limits~\cite{Munley2024}.
    
    Another line of future research is how to deal with RAG data dependence of the quality and relevance of the data used as input (e.g., retrieval sources), a topic discussed in general~\cite{Barnett2024} and also in testing-specific studies~\cite{Chetan2024}. 
    Finally, some authors have also highlighted the challenges of RAG to capture the necessary information for testing~\cite{Munley2024}.
    
    To address those challenges, some authors have proposed the inclusion of a \textit{human-in-the-loop} to supervise and refine the RAG pipeline~\cite{Chetan2024}. Others have pointed out room for improvement, finding that RAG alternatives~\cite{Khandaker2025} are outperformed by the SOTA tools.
    
    \begin{acmmdframed}
        \textit{How to retrieve relevant and accurate information, aligned with the testing objectives, remains the major challenge in RAG usage. Future directions involve the quality and quantity of the information, as well as tailoring its content to the testing needs.}
    \end{acmmdframed}

\end{itemize}

\subsubsection{Validation}
\label{sec:validation}

Evaluating how the LLMs perform is an open challenge acknowledged by several works~\cite{Alshahwan2024, Augusto2025, Chen2023, Hailong2024}. Among the different challenges, we stand out two: (1) the hallucination phenomenon, which can hinder output quality, and (2) the lack of robust, domain-specific evaluation metrics. These challenges have a direct impact on the correctness, trustworthiness, and reproducibility of LLM-based testing, as discussed below:

\begin{itemize}
    \item \textbf{Hallucinations:} The major challenge of LLM-based testing is addressing the phenomenon of hallucinations. Several authors have acknowledged this as a critical issue~\cite{Fraser2025, ChenArxiv2024, SuYanqi2024, Kang2024, LiuISSRE2024, Le2024, Yang2025, Bianou2024}, particularly due to its impact on the correctness and validity of the generated outputs. When hallucinations are produced, they often result in flawed code~\cite{Plein2024, Vikram2024, Yuan2024, WangArxiv2024, Bhatia2024, Lops2024, Watson2020, Primbs2025, Yu2023, Zheng2025, Munley2024}, or inexact outputs/calculations~\cite{Karmarkar2024}. \textit{This leads to substantial concerns about reliability, correctness, and trustworthiness when using LLMs for testing.}
    
    However, hallucinations are not perceived negatively in all testing fields. An emerging body of research has started to consider hallucinations positively. Specific authors highlight the capacity of LLMs to perform as \textit{``out-of-the-box''} testing thinkers of generating non-obvious configurations, inputs, or strategies that escape from the common heuristic and can be valuable, e.g., in test exploration~\cite{XueISSTA2024, Karmarkar2024}. 
    \textit{This perspective suggests that properly harnessing the hallucinations may extend the creative boundaries of testing by offering diverse inputs and scenarios that expose edge cases that the test engineer might overlook.}
    
    Still, the risks of uncontrolled hallucinations require effective mitigation strategies. To this end, some authors recognize a line of future work that introspects the model by analyzing token-level probabilities as an indicator of output reliability~\cite{Taherkhani2024}. Complementary, other approaches have presented post-processing filters that discard those invalid outputs generated by the hallucinations~\cite{Almutawa2024}.

    \begin{acmmdframed}
        \textit{The dual role of hallucinations as both a reliability problem and a source of creativity presents a rich avenue for research. Addressing them with strategies that not only remove hallucinations but also understand and channel them would be beneficial for some applications in the testing life cycle.}
    \end{acmmdframed}

    \item \textbf{Metrics:} The assessment of LLM-based testing approaches remains an open challenge, despite increasing attention from practitioners and the research community. Several authors have explicitly acknowledged the lack of robust, effective evaluation metrics~\cite{Tufano2022, Augusto2025}. The problems faced depend on the nature of the LLM-generated output. For instance, when the output is in natural language,  the existing NLP domain metrics such as BLEU, ROUGE, and METEOR have been successfully applied to measure the syntactic similarity between LLM and the human tester output.
    However, these metrics were initially used to evaluate code generation, revealing several limitations:  they often produce misleading evaluations by failing to capture functional equivalence between syntactically distinct and semantically correct code segments~\cite{Hailong2024, evtikhiev2023out}. As a counter response, the research community has proposed more nuanced and domain-specific metrics that align with the human programmer's sense of correctness and utility. Emerging proposals include CodeBLUE, Match Success Rate (MSR), Code Extraction Rate (CSR)~\cite{OuedraogoArxiv2024}, as well as more granular criteria such as post condition completeness and correctness~\cite{Madeline2024}.

    \begin{acmmdframed}
        \textit{ The evaluation metrics for LLM-based testing face technical and epistemological challenges: defining what is ``good'' for humans and tools and measuring them consistently and effectively. With the adoption of LLMs in testing, this gap becomes a foundational step for validation, trust, and further progress.}
    \end{acmmdframed}
  
\end{itemize}
\subsubsection{Social Challenges}
\label{sec:social}
Beyond technical challenges, adopting LLMs in software testing has several human-centric challenges. Among these challenges, we have identified the trust in the models, the economic, technical, or performance entrance barriers for the practitioners, and the need for educational frameworks to ensure responsible and effective LLM use. These challenges, which are less visible but play a key role in the future of LLM-based testing, are covered and discussed in the \textit{trust}, \textit{adoption}, and \textit{education} sections below:

\begin{itemize}
    \item \textbf{Trust:} It remains a big challenge in the applicability and adoption of  LLMs in software testing. The root of the problem is the common ambiguity~\cite{Liu2025}: \textit{``when the test fails, is it due to a genuine bug in the PUT, or is it a failure of the model?''}.
    Since LLMs are trained on human-produced data, these models are prone to replicating not only best practices, but also common mistakes and code anti-patterns~\cite{Alshahwan2024, OuedraogoArxiv2024}. Many studies have raised concerns about the correctness of the generated outputs, such as code snippets, test cases, or mutants~\cite{Plein2024, Vikram2024, Yuan2024, WangArxiv2024, Bhatia2024, Lops2024, Watson2020, Primbs2025, Yu2023, Zheng2025, Munley2024}, alongside the accuracy in the calculations~\cite{Karmarkar2024}.
    Several open directions have been proposed to mitigate these issues and reinforce trust. One direction focuses on enriching the model's input with more context, e.g., providing the error feedback to guide the generation process~\cite{Lops2024}.
    Another involves hybrid systems integrating LLMs with more reliable SOTA tools~\cite{Dakhel2024}. Further proposals include enhancing explainability, simplifying and clarifying the complex generated outputs~\cite{Rahman2024}, even chaining multiple models in a pipeline to refine the output iteratively~\cite{Almutawa2024}.
    
    Another factor undermining trust is that LLMs often struggle to perform well with complex or specific scenarios requiring advanced reasoning or semantic understanding.~\cite{Myeongsoo2024, Hayet2025, SuYanqi2024, Nie2023, Yibo2024, ZhangTSE2025, Yifan2023, Zhang2025}. Examples of those complex scenarios are generating oracles that handle exceptions~\cite{Khandaker2025}, generating multi-assert test cases~\cite{Watson2020} or the full program state~\cite{Alian2025}, generating complex mutations~\cite{ZhangAST2025}, or recognizing subtle, system-level faults~\cite{Myeongsoo2024}.
    Many researchers propose introducing the \textit{“human-in-the-loop”} to bridge this gap. By involving testers to guide, inspect, or augment the model's decisions, this approach enhances the trustworthiness and makes the model consider situations that may be overlooked~\cite{Chetan2024}. The tester can validate outputs~\cite{Augusto2025}, introduce missing domain knowledge~\cite{PaduraruJournal2024, TsigkanosICCS2024}, and even increase the creativity of the inputs to complement the model capabilities~\cite{SuYanqi2024}.

    Finally, and unsurprisingly, trust itself also relies on social aspects. Thus, within the community of developers using LLMs, trust can be fostered~\cite{cheng2024would} through collective sense-making, where developers build trust in others' experiences and shared opinions (e.g., voting).

    \begin{acmmdframed}
        \textit{While LLMs hold potential for automated testing, trust is a barrier due to ambiguous failure interpretations, error-prone outputs, and difficulties with complex scenarios. Enhancing collaboration between humans and reliable SOTA tools can boost confidence in LLM-generated results.}
    \end{acmmdframed}

    \item  \textbf{Adoption:} the integration of LLMs into all software testing processes is promising but far from seamless. Adoption faces four challenges: performance, cost, privacy, and usability.
    The first is the performance challenge, which has been mentioned as a relevant factor of many of the above-described challenges. While LLMs have demonstrated potential, their application often is time-consuming~\cite{Hayet2025, Yang2025}, resource-inefficient (e.g., frequent API calls, or a lot of generated outputs since the correct is achieved)~\cite{Altmayer2025, Fraser2025, Pan2025}, or requires a lot of manual post-processing~\cite{LiDTPI2024, Augusto2025}.
    Several studies show that LLMs in isolation frequently outperform SOTA tools crafted explicitly for this purpose (e.g., improve coverage)~\cite{Ryan2024, Siddiq2024}.
    This has led to increasing interest in combining LLMs with traditional SOTA tools to leverage the strengths of both~\cite{Primbs2025, Fraser2025, Konstantinou2024, SuYanqi2024, SantosSBQS2024, Kang2023, Petrovic2024, XiaoInt2024}.
    However, such integration is still negatively affected by the lack of plugins and tool support in some cases~\cite{Feldt2023}.
    The second barrier is cost and infrastructure. Running LLMs locally or fine-tuning them requires expensive and specialized hardware, and LLMs-as-a-service can also be costly or limited by data regulations policies.
    This makes it not feasible for most start-ups and small to medium organizations to leverage the potential of these models~\cite{Feldt2023, Tufano2021}.
    Third, privacy concerns, already mentioned, remain significant. Using LLMs via external services raises risks of data leakage and unclear usage policies~\cite{Feldt2023}, hindering their adoption, particularly in regulated sectors that handle sensitive data.
    Finally, the fourth challenge relates to usability.
    The adoption of LLMs implies dealing with two aspects. The former is about seamless technical integration with the existing development workflows~\cite{Russo2024}. The latter requires a better understanding of human–AI interaction, getting insights into how testers interpret and utilize LLM outputs~\cite{Shi2024}.
    
    \begin{acmmdframed}
    \textit{Addressing the challenges of adopting LLMs testing requires coordinated efforts in performance tuning, tool support, cost reduction, privacy protections, technical integration, and human-AI collaboration.}
    \end{acmmdframed}

    \item \textbf{Education:} The responsible and effective use of LLMs in software testing requires integrating them into education and training programs~\cite{Santos2024}. 
    Employing LLMs for testing can be intellectually challenging and requires specific knowledge and prompting skills~\cite{XueISSTA2024}, not included in the current learning roadmaps. 
    A further obstacle is the lack of standardization, which is starting to be addressed through the development of new taxonomies~\cite{Braberman2024}.

    \begin{acmmdframed}
    \textit{In sum, bridging the educational gap is essential to scale the practical use of LLMs in testing. This will require coordinated efforts in curriculum design, tool development, and standardization.}
    \end{acmmdframed}
\end{itemize}

\subsection{How are LLMs going to impact software testing research}
\label{sec:reflections:dreams}

\begin{figure}[ht]
\centering
\includegraphics[width=0.9\textwidth,trim=100 40 100 40, clip]{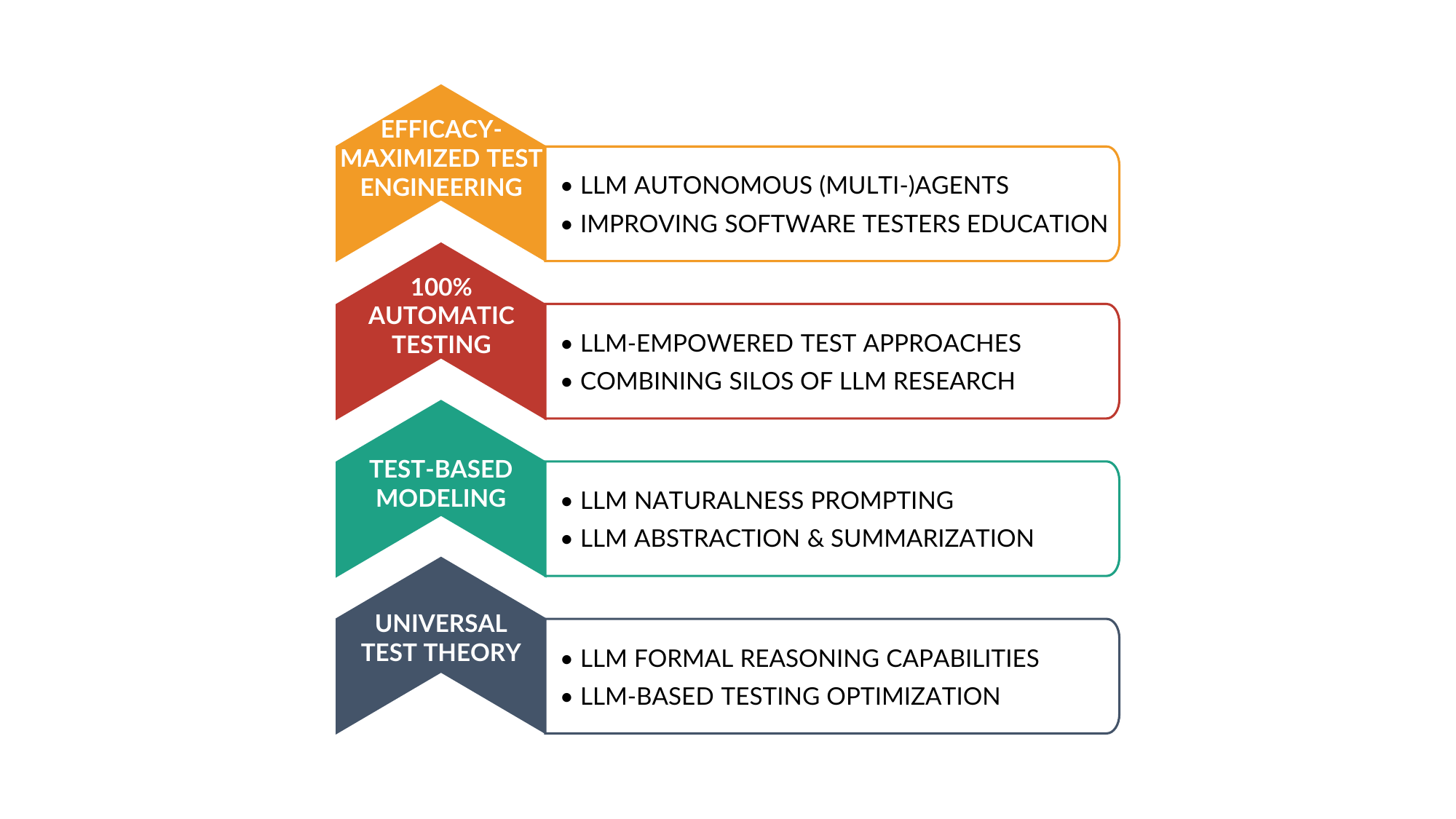}
\caption{LLM-Software Testing Impact on Dreams} \label{fig:reflecimpact}
\Description{LLM-Software Testing Impact on Dreams}
\label{fig:dreams}
\end{figure}

Beyond the above-identified challenges, the current momentum that LLMs have is affecting the whole research in software testing.  In this section, we attempt a perspective of such an impact, and to do so, we leverage the concepts already introduced in a previous highly-cited roadmap on software testing research~\cite{fose2007}.

Specifically, that roadmap projects achievements and challenges in software testing research towards four \textit{dreams}, defined as the final destinations to which research ``\textit{ultimately tends, but which remain as unreachable as dreams}''~\cite{fose2007}. In fact, although that work is almost 20 years old, and certainly some of the achievements/challenges it discusses would need to be revised in light of more recent results, the four dreams (faithfully to their definition) still appear actual and relevant. Thus, in this work we propose our prospect of whether and how LLMs can foster software testing research to get closer to its dreams.

According to~\cite{fose2007}, the four software testing dreams are related in a sort of hierarchical utility (see on the left-hand side of Fig.~\ref{fig:dreams} can be leveraged to approach the second dream of \textit{Test-based modeling}; proper models allow for pushing automation, towards the third dream of \textit{100\% automatic testing}. 
Finally, automation is instrumental to a cost-effective test process towards the topmost dream of \textit{Efficacy-maximized test engineering}.
The rest of the section discusses in detail how LLMs could contribute to these dreams.

\subsubsection{Universal test theory}
Testing is a pragmatic activity, whose industrial application still has to solve complex challenges and lacks a general theory~\cite{garousi2020exploring, rafi2012}. The vision of the roadmap is that assumptions, limitations and capabilities of test techniques and tools could be collected into a \textit{coherent and rigorous framework}~\cite{fose2007} that informs and guides testers in choosing, composing and applying the test approaches that are the most adequate to their situation, thereof reducing the dependency of success on tester's skills and background.

From our survey of current research in LLM-based testing, we did not find so far works explicitly dealing with leveraging the LLM potential to build such a theoretical framework: yes, we can mention several studies that focus on the limitations and potential of LLM-based approaches themselves, mostly through empirical studies. 
For example, both Ou\'edraogo et al.~\cite{OuedraogoArxiv2024} and Li et al.~\cite{Li2025} conduct in-depth analyses of empirically evaluating and comparing different LLMs using various types of prompts and contextual information. The former compares the results to the EvoSuite baseline, while the latter compares them against manual testing. Both studies provide insightful reflections about the respective strengths and weaknesses of the analyzed models.

There also exist few studies that analyze the LLM-based testing literature,  aiming at proposing a rationalization of what LLMs can achieve. Notably, Braberman et al.~\cite{Braberman2024} review LLM-enabled architectures used for software testing and analysis, and propose an extensive taxonomy of \textit{``downstream tasks''} achievable by LLM prompting.

Such empirical or analytical studies are certainly enlightening, but this is not what we are interested in here: we ask if LLMs could support an analysis of existing techniques and tools to help understand and codify their mutual applicability, composability, and achievable results. Even though this potential use of LLMs has not been explored yet, we are anyhow optimistic that in future the LLMs formal capability to analyze and relate facts could reveal itself as a useful tool towards getting closer to this dream. 

We foresee that LLM-based testing will indeed act as both an indirect and a direct lever to strengthen the theoretical foundations of software testing. Indirect because in most works we overviewed, a conscientious effort can be observed to find the best approach and the most effective information for prompting the LLM; for example, the prompt generation method for test case generation by Gao et al.~\cite{Gao2025} includes a module for the extraction of domain contextual knowledge that helps optimize the accuracy of test cases. 
Said in different words, research focused on optimization of LLM-based testing requires that we better understand the essential and correct information needed for a testing activity, and this is clearly very relevant for the formulation of the universal test theory we dream of. 

As a direct lever, 
we forecast that the powerful reasoning capabilities of LLMs (as advocated in~\cite{yang2024formal}), on top of their power to process and synthesize huge amounts of data, could be used for the generation of test hypotheses, one of the main challenges to address for understanding the actual effectiveness of a test criterion~\cite{fose2007}.
As said, we did not find works in this domain, but the very idea of leveraging LLMs for hypotheses generation is already used in other domains, see, e.g.,~\cite{zhou2024hypothesis}.

\begin{acmmdframed}
\textit{No impact of current LLM-based testing research, but in the future it could help to lay the foundations of a universal test theory both indirectly, e.g.,  by studying the most effective prompts, and directly thanks to LLM formal reasoning capabilities, e.g., to derive test hypotheses.}
\end{acmmdframed}

\subsubsection{Test-based modelling}
In this dream, the software to be tested comes equipped with a model conceived to facilitate testing, carrying in a well-structured way all information needed for testing purposes. This would include provided functionality, assumptions about the execution environment, possible input constraints, the expected behavior for each given input, etc.

LLMs can certainly push progress towards this dream: on the one side, several works leverage LLMs abstraction and summarization capabilities to derive a (formal) specification for software, e.g.,~\cite{Endres2024, Ma2025} among others; on the other side, as we presented in Section~\ref{sec:llmusage:oraclegen}, much research has already been devoted to the derivation of test oracles, which in the roadmap is presented as one of the hardest challenges that has to be addressed for achieving this dream.

Besides, LLMs have demonstrated they can yield great creativity~\cite{castelo2024ai}, thanks to which they could also be leveraged to provide highly expressive and human-understandable models, by combining textual, graphical, and other forms of communication, into a multi-modal (and more natural) specification~\cite{wang2024multi} of the software to be tested. 

However, looking ahead, the perspective seems that LLMs are likely to replace human testers for the tasks of test case generation, execution, and validation, while the task of testers will be that of prompting the LLM and checking its answers. In this view, we foresee that the dream of test-based modelling will lose relevance, given the capability of LLMs to process NL or informal inputs. If we add to this the related possibility that the code to be tested is generated by LLMs as well, the utility/necessity of intermediate artifacts is going to decrease.

\begin{acmmdframed}
\textit{LLMs can facilitate the redaction of better models for testing purposes, by synthesizing formal or graphical specifications of a function behaviour, as well as by deriving test oracles. In perspective, though, it is likely that usage of LLMs will make model artifacts less important.}
\end{acmmdframed}

\subsubsection{100\% automatic testing}
The dream is that of a powerful integrated test environment, which monitors the progresses of software development or maintenance to detect the need for testing, and is able to autonomously manage the setting, configuration, execution, reporting of testing, as well as bug localization and repair where needed.

It is evident that LLMs are already impacting the pursuing of this dream in each and every testing activity: indeed, pushing automation beyond the levels achieved by existing tools is the ultimate motivation for using LLMs. 
We have reviewed in Section~\ref{sec:llmUsage} 
many works presenting tools for automating test activities: in particular, LLMs support test generation, both at unit and at system levels, as we discussed in Sections~\ref{sec:llmusage:unittestgen} and~\ref{sec:llmusage:highlevel}, respectively; other works rely on LLM-empowered agents for test configuration or execution, as we overviewed in Section~\ref{sec:llmusage:testagents}; LLM-based approaches also facilitate the derivation of automated oracles, as we illustrated in Section~\ref{sec:llmusage:oraclegen}. In perspective, the ongoing research shows that LLMs will certainly boost software testing automation. 

However, research also warns that, while improving test automation, LLMs are also introducing new problems and new challenges, which require careful attention. We have highlighted several open challenges in the previous section: also when used in software testing, LLMs can suffer from hallucinations, thus LLM responses need appropriate assessment and validation metrics. Moreover, several studies have shown that LLMs' performance depends heavily on the prompt provided by testers, so that human efforts that can be saved thanks to test automation should be redirected towards the prompt engineering stage. 

As a final important observation, in our review of literature we also noticed that research in LLM-based testing is proceeding within separated silos, but eventually software testers need to automate the \textit{whole process} by combining all different activities, and not the individual activities in separation: in fact this dream envisions an \textit{integrated test environment}, but we did not find efforts towards integrating differing activities so far.

\begin{acmmdframed}
\textit{LLMs are already pushing automatic testing, but at the same time, they are also bringing new challenges, and are requiring testers to change their expertise. Moreover, so far research is proceeding within separate silos, and we are far from getting closer to an integrated LLM-based test environment.}
\end{acmmdframed}

\subsubsection{Efficacy-maximized test engineering}
This is proposed as the very ultimate goal of software testing research, and is defined as the ideal -yet at same time practical- objective of engineering viable methods, processes, and tools apt to develop high-quality software in cost-effective way.
The ``efficacy'' term in the dream name is used to denote both properties of efficiency and effectiveness~\cite{fose2007}.

Strictly speaking LLMs have not been leveraged so far to assess or improve the cost-effectiveness of testing activities, in the sense that we did not find in the reviewed literature an explicit concern about using LLMs for reducing the high costs of testing. 
Several works use LLMs for improving test coverage, e.g.,~\cite{ChenFSE2024, WangTESTEVAL2024, Altmayer2025} among many others; however, as discussed by Mathews and Nagappan~\cite{Mathews2024}, current solutions could not be well thought towards improving failure detection effectiveness.
Of course, there exist various LLM-based approaches that show an improvement of fault-detection effectiveness with respect to a traditional baseline: for instance~\cite{OuedraogoASE2024, Liu2024, Zhong2024}.
However, some authors also warn about the high costs implied by the use of LLMs, e.g.,~\cite{ChenArxiv2024}.

In perspective, we are confident that once LLM-based approaches to test generation and execution acquire higher maturity, LLMs could  provide good support towards addressing some of the challenges earlier identified in Bertolino's roadmap~\cite{fose2007}. In particular, LLM-empowered agents could support adaptive and autonomous testing of evolving software, e.g.,~\cite{Bouzenia2024, Yoon2024}, thus relieving testers from the burden of understanding when and what to retest as software evolves. Moreover, it is evident the potential influence that LLMs can have in the education of software testers, which was also indicated in~\cite{fose2007} as a promising direction to pursue towards a more cost-effective test process.

In the long term, we somehow could assist to the reality surpassing this dream, if we will be able to make significant progress in the development of multi-agent solutions (as in~\cite{Garlapati2024}) that can monitor the software in execution, launch test cases (perhaps also in the field~\cite{BertolinoField2021}), and even autonomously repair the software when faults are detected. However, again, we need also to take into account the involved costs, pushing research in leveraging LLMs to assess direct and implied costs of software testing.

\begin{acmmdframed}
\textit{LLMs have not been leveraged to evaluate or reduce the costs of testing, and in some cases, their use has rather augmented those costs. However, in perspective, we expect huge potential of LLMs towards improving the cost-effectiveness of the testing process, either by means of multi-agent frameworks, or by improving testers' education.}
\end{acmmdframed}

%% file: 07_conclusion.tex
\section{Concluding remarks}
\label{sec:conclusion}

Triggered by the recent explosion of interest into LLMs and, more specifically, into their growing use to support software testing activities, in this work we aimed at bringing order into research in LLM-based testing. 
This study is not meant as a systematic literature review; rather we propose a roadmap of current research results and open challenges, which is built on top of a semi-systematic review.
We collected ongoing research under seven categories, namely: Unit Test Generation, High-Level Test Generation, Oracle Generation, Test Augmentation or Improvement, Non-Functional Testing, Test Agents, and Reflections. For each category (except for the works in the Reflections category, which inspired instead our roadmap directions), we provided a generic scheme of the underlying process and an overview of ongoing research. 
The final product of our study consists of a discussion highlighting open challenges and envisioning the possible impact of LLMs on the software testing research discipline. 

We also discuss whether and how LLMs can help software testing researchers get closer to their software testing dreams. 
LLMs represent an effective support for testing activities and are frequently integrated with, or employed alongside, traditional SOTA tools. Although their adoption can yield substantial cost and resource savings, these benefits must be weighed against the additional effort required for complementary activities—such as prompt engineering and fine-tuning—necessary to assure high test quality.
Many challenges related to automation, reliability of LLM output, social aspects, and education need to be still carefully addressed.

While many open problems still face us, as we discuss in the article, we trust that the insurgence of LLMs and Generative AI in software testing is not a temporary wave, but a revolution that is going to produce permanent effects. We acknowledge that the approaches and tools for LLM-based testing are still far from mature, and indeed expect huge progress to come in the very near future. Nevertheless, we got convinced that LLMs' capabilities of analysis and summarization of NL and code will provide developers with reliable support for autonomous, systematic, and rigorous verification: note that we claim this in perspective and not with reference to the current state of play, which is still in evolution.
Several lines of future work can be sketched from the findings and open challenges highlighted in this article, aligned with the researcher's expertise and research lines, e.g., system testing, reinforcement learning testing, test generation, or Cloud Testing. 
We hope that our roadmap can be helpful to researchers for better directing their future investigation in LLM-based testing. Conversely, we believe that continuous monitoring of emerging works can provide a validation and possible revisions of our roadmap.